\documentclass[12pt]{article} 

\title{Spin 1/2 and Invariant Coefficients II. Massless }  
\author{{\it Richard Shurtleff~}\thanks{affiliation and mailing 
address: Department of Applied Mathematics and Sciences, 
Wentworth Institute of Technology, 550 Huntington Avenue, 
Boston, MA, USA, ZIP 02115, telephone number: (617) 989-4338, fax 
number: (617) 989-4591 , e-mail addresses: shurtleffr@wit.edu; momentummatrix@yahoo.com (preferred), download associated Mathematica notebook at ox.wit.edu/~shurtleffr}} 
\begin{document} 

\maketitle 
%\pagebreak

\begin{abstract} 

A `covariant' field that transforms like a relativistic field operator is required to be a linear combination of `canonical' fields that transform like annihilation and creation operators and with invariant coefficients. The Invariant Coefficient Hypothesis contends that this familiar construction {\it{by itself}} yields useful results. Thus, just the transformation properties are considered here, not the specific properties of annihilation or creation operators. The results include Weyl wave equations for some massless fields and, for other fields, Weyl-like noncovariant wave equations that are allowed here because no assumptions are made to exclude them. The hypothesis produces wave equations for translation-matrix-{\it{invariant}} fields while translation-matrix-{\it{dependent}} coefficient functions have currents that are the vector potentials of the coefficient functions  of those translation-matrix-invariant fields. The statement is proven by showing that Maxwell equations are satisfied, though in keeping with the hypothesis they are not assumed to hold. The underlying mechanism is the same for the massless class here as it is for the massive class in a previous paper, suggesting that spin 1/2 particles may have a universal electromagnetic-type charge whether they are massive or massless.

\end{abstract}

\pagebreak
          
\section{Introduction}

Successive infinitesimal rotations, boosts, and translations transform spacetime yet preserve the spacetime metric. Among the many ways to represent these `inhomogeneous Lorentz' or `Poincar\'{e}' transformations are the two employed here, the covariant and the canonical representations. 

The Invariant Coefficient Hypothesis is the idea that useful constraints on fields can be obtained by requiring a covariant vector field $\psi$ to be constructed as a linear combination of canonical vector fields $a$ with invariant coefficients $u,$ $\psi$ = $\sum u a.$ Considerations motivating such a construction can include the contrast between the principles of quantum mechanics that require unitary canonical particle states and the need to have covariant fields to make S-matrices. So it is natural to build useful covariant fields from the fundamental canonical fields.\cite{Weinberg} 

The Invariant Coefficient Hypothesis takes a somewhat different approach by removing all assumptions based on quantum or relativistic principles. Obviously such restrictions can be made when applying the results found here, but for a cleanly stated mathematical problem extraneous assumptions are avoided. The canonical representations are not assumed at the outset to be unitary and the non-covariant wave equations derived below are not discarded.

The covariant vector field $\psi_{l}(x)$ has a discrete index $l$ together with continuous parameter space of `coordinates' $x^{\mu}$  = $\{x,y,z,t\},$ $\mu \in$ $\{1,2,3,4\}.$ A Poincar\'{e} transformation of the field is in part a differential representation that changes the function of $x$ to a function of $\Lambda x +b,$ where $\Lambda $ is a homogeneous Lorentz transformation and  $b$ is the displacement. The discrete index $l$ exists to be transformed by a matrix, $\psi_{l} \rightarrow$ $D^{-1}_{l \bar{l}}(\Lambda,b)\psi_{\bar{l}},$ where the summation convention is in force and the collection of matrices $D^{-1}(\Lambda,b)$ form a spin 1/2, nonunitary, finite dimensional representation of the Poincar\'{e} group. The transformation matrices $D(\Lambda,b)$ are constant, independent of the continuous parameters $x^{\mu},$ which is a useful feature in some applications.

In contrast, the canonical representations have matrices that depend on values in a continuous parameter space labeled $p^{\mu}.$ The canonical vector fields are actually a collection of vector fields from which a vector field with a canonical transformation representation compatible with the covariant representation is selected. The canonical vector field $a({\mathbf{p}},\sigma)$ has a discrete index $\sigma$ together with a continuous parameter space called the `momentum' $p^{\mu}.$ Since the transformation matrix ${D^{({\mathrm{Can}})}}^{-1}_{\sigma \bar{\sigma}}(W(\Lambda,p))$ depends on a continuous parameter $p^{\mu},$ the set of finite dimensional matrices forms an infinite dimensional representation of spacetime transformations. Unitary representations of non-compact groups such as the Poincar\'{e} group must be infinite dimensional and canonical representations can be unitary. But they need not be unitary and are not assumed to be unitary in this article.

Poincar\'{e} transformations separate into classes distinguished by inequivalent Wigner little groups.\cite{Wigner} The Wigner little group of a special  4-vector consists of the transformations $W$ that preserve the special 4-vector as well as preserving the metric.  The expression $W(\Lambda,p)$ in the canonical transformation matrix ${D^{({\mathrm{Can}})}}$ above indicates that  each transformation $\Lambda$ determines a little group transformation $W$ for each momentum $p^{\mu}.$

A previous article, paper I,\cite{paperI} dealt with the spin 1/2 time-like, massive class for which momenta are limited by $p_{\mu} p^{\mu}$ = $M^2,$ where $M$ is the mass. The special 4-vector for a massive particle can be its momentum at rest, $p^{\mu}$ = $\{0,0,0,M\}.$ Clearly rotations form the little group for massive particles since rotations leave the time component unchanged and rotating null spacial components is a futile effort, producing null spacial components. Thus the possible canonical representations for the massive class are the collection of representations of rotations. 

By the hypothesis, the canonical representation must be compatible with the covariant representation. For the massive class this means the covariant representation of rotations must be equivalent to the canonical representation. Since the covariant spin 1/2 representation of {\it{rotations}} for spin $(0,1/2)$ and for $(1/2,0)$ are (i) equivalent and (ii) unitary, there is (i) a single canonical representation which must be (ii) unitary. Neither characteristic holds for the massless class.

In this paper, the hypothesis is applied to the spin 1/2 massless class of Poincar\'{e} transformations with momenta limited to those with $p_{\mu} p^{\mu}$ = $0,$ with $p^{t} >$ 0. The special 4-vector, here taken to be $k^{\mu}$ = $\{0,0,k,k\}$ with $k >$ 0,  has a little group $W$ that consists of transformations involving both rotations and boosts, as is well known.\cite{T} The boosts have characteristics that differ from rotations.

 By the hypothesis, the canonical representation must be compatible with the covariant representation. For the massless class this means the covariant representation of the little group $W$ must be equivalent to the canonical representation. But boost generators in the $(0,1/2)$ and $(1/2,0)$ representations are not equivalent and not unitary. Since the  transformations $W$ involve boosts, the covariant spin 1/2 representation of $W$ for spin $(0,1/2)$ and for $(1/2,0)$ are (i) not equivalent and (ii) not unitary. Thus (i) there are at least two canonical representations and (ii) no representation is unitary.

In fact, there is an exception to characteristics (i) and (ii) that occurs for a one dimensional unfaithful covariant representation of the massless little group $W.$ Essentially, the transformations $W$ that involve boosts are represented trivially by the unit matrix, so the remaining $W$s are rotations, for which the canonical representation is unique. Restricting discussions to the unfaithful representations is common in the literature.\cite{T,W,vdBij} Sometimes it is argued that nonunitary representations are allowed if they can be restricted to gauge transformations since gauges are unobservable.\cite{Han,K} Such considerations are beyond the scope of this article and may be explored elsewhere.

In this paper the unfaithful case is just one of three possible choices for the canonical representation. Let choice $I$ involve a canonical representation  equivalent to the $(0,1/2)$ representation of the little group $W$ and let $II$ indicate the representation is to be equivalent to the $(1/2,0)$ representation of $W.$ Finally let $III$ represent the unfaithful case. Each step of the work must be done three times. The three canonical transformation characters are labeled $A \in$ $\{I,II,III\}$ for those occasions when the three cases give expressions that can be written in common. 

But then there are the adjoint representations: to each canonical representation $A$ corresponds an adjoint representation $B \in$ $\{IV,V,VI\}$ = $\{I^{\dagger},II^{\dagger},III^{\dagger}\} .$ So there are six canonical transformation characters to consider, each with canonical vector fields $a^{(A)}$ or $a^{(B)\dagger},$ with which to construct a covariant field $\psi^{(A)}$ or $\psi^{(B)}.$ In the end there are just three distinct transformation characters because the adjoints are similar to the original three,  $\{IV,V,VI\} \simeq $ $\{II,I,III\} ;$ representations $I$ and $II$ are each other's adjoint while $III$ is its own adjoint. Indeed, there can only be three inequivalent canonical representations because there are just three inequivalent covariant representations. The adjoint representations are included to accommodate applications that involve positive and negative energy states.

\begin{table}
\centerline{
\begin{tabular}{|c||c|c|c|} \hline
{\it Spin} & {\it Block Matrix:} & {\it Handedness} & {\it Coefficient Function:} \\ 
  &  $\pmatrix{11&& 12 \cr 21&& 22}$  &  & $\pmatrix{u_{+} \cr u_{-}}$ \\
  \hline \hline
$(0,1/2)$ & 11 & Right  & $u_{+}$ \\ \hline
$(1/2,0)$\ & 22 & Left  &  $u_{-}$ \\ \hline
\end{tabular}}
\caption{Alternate Indicators of Spin. The spin discussed in this paper is $(0,1/2)\oplus (1/2,0),$ which has angular momentum matrices that can be represented in block diagonal form, see paper I, equation I(7). Each block is a $2 \times 2$ matrix. The terms Right- and Left-handed indicate spin $(0,1/2)$ or $(1/2,0)$ especially when discussing spatial inversions and parity. } The $u_{+}$ and $u_{-}$ parts of a coefficient function are each a 2-component column matrix.
\end{table}

Turn now to translations. A translation along 4-vector $b$ changes the covariant field $\psi_{l}(x)$ in two ways, the field's components are mixed via matrix multiplication and the function of position changes, $\psi_{l}(x) \rightarrow$ $D^{-1}(1,b)\psi(x + b),$ where $D(1,b)$ is the translation matrix.  While all fields depend on translations because they are functions of position, the fields neatly split into translation-matrix-{\it{invariant}} and translation-matrix-{\it{dependent}} fields.

As discussed more fully in paper I, there are two translation matrix representations, labeled `$12$' and `$21$' which indicates their nonzero blocks when displayed with a particular choice of $\gamma$ matrices. Considering just the canonical representations $A,$ that makes six combinations: two covariant representations $12$ and $21$ with the three canonical representations $A.$  One finds that some combinations of $A$ with $12$ and $21$ are translation-matrix-invariant, i.e. $\{(12,I),(21,II)\},$ while the fields with labels $\{(21,I),(12,II)\}$ are translation-matrix dependent. Type $III$  fields split similarly, but in a way that depends also on helicity $\sigma.$ 

Translation-matrix-{\it{invariant}} coefficient functions depend on position just in the momentum dependent phase factors $\exp{(\pm i p\cdot x)};$ the fields are sums of plane waves. A consequence of the Invariant Coefficient Hypothesis is that translation-matrix-invariant fields obey simple wave equations. These first order partial differential equations are of one of two forms, either obviously covariant, $\eta^{\mu\nu} \gamma_{\mu} \partial_{\nu} \psi(x)$ = 0, or not covariant, $\delta^{\mu \nu} \gamma_{\mu} \partial_{\nu} \psi(x)$ = 0, where $\eta^{\mu\nu}$ is the metric, here taken to be diag$\{-1,-1,-1,+1\},$ and $\delta^{\mu\nu}$ is the Kronecker delta function, effectively a positive definite metric. Thus both non-covariant wave equations and covariant wave equations follow from the Invariant Coefficient Hypothesis.

The {\it{noncovariant}} wave equations arise from equations for coefficient functions that are {\it{covariant}}. Since the field is a sum of plane waves proportional to $\exp{(\pm i p\cdot x)}$, and the momentum $p_{\mu}$ results when the `momentum operator' $i \partial_{\mu}$ is applied to the phase factor, to get a wave equation each of these equations is rewritten in terms of the momentum $p_{\mu}.$  Each equation for the coefficient function is covariant, but to get an equation with just  $p_{\mu},$ a covariant-preserving factor cancels, thereby introducing noncovariance. Coefficient-by-coefficient the equations are covariant, but collected together the same equations give a noncovariant wave equation.

The translation-matrix-{\it{invariant}} covariant fields constructed from type $I,II$ canonical vector fields have definite parity. These fields transform by either the $(0,1/2)$ right-handed or the $(1/2,0)$ left-handed spin 1/2 representations. See Table 1 for the terminology. As a consequence of Schur's Lemma the left-handed part of the right-hand transforming field must vanish and visa versa. Translation matrices would mix left and right-handed parts of the fields, but these fields are precisely those fields that are unaffected by translation matrix multiplication and have definite parity. For example a massless, left-handed, positive-energy, spin 1/2 particle might be described by the covariant fields $\psi^{(21;II)}(x)_{\sigma=+1/2}$ and $\psi^{(21;II)}(x)_{\sigma=-1/2}.$ The $\sigma=+1/2$ field obeys a non-covariant wave equation; while the $\sigma=-1/2$ field obeys a covariant wave equation. 

Translation-matrix-{\it{dependent}} coefficient functions form currents that depend on position $x^{\mu}$ in either the matrix product $D^{(12)\dagger}(1,x)$ $\gamma^{4}\gamma^{\mu}D^{(12)}(1,x)$ or its 21-version. These same expressions appear with both the massive class and the massless class. They are quadratic in $x^{\mu},$ so their second partial derivatives with respect to  $x^{\mu}$ are constant. It happens that the second partials are just what is needed for the translation-matrix-{\it{dependent}} coefficient current to be the vector potential for the constant current of the translation-matrix-{\it{invariant}} coefficient, automatically.

Since Maxwell equations occur here with the massless class via the same mechanism as it does in the massive particle class, the Invariant Coefficient Hypothesis suggests that an electromagnetism-type interaction is universal for spin 1/2 covariant fields. However, interactions also involve a response to force fields such as a Lorentz force law and that aspect of interaction is not discussed in this article or in paper I.

Much of the spin 1/2 formalism needed for this article can be found in paper I. Section \ref{Poincare} updates the spin 1/2 formalism for the  little group  of the special light-like 4-vector $k^{\mu}$ = $\{0,0,k,k\}$ with $k >$ 0. The Invariant Coefficient Hypothesis in Section \ref{derive} is applied to the construction of three covariant fields that are linear combinations of three sets of canonical field vectors with different transformation characters. The coefficient functions for each case are determined aside from a normalization constant. Section \ref{1221field} shows that fields $\psi^{(12;A)}$ and $\psi^{(21;A)}$ are not related by the 12/21 transition, a parity relationship. Parity considerations for massless particles naturally differ from those for the massive particle class. In Section \ref{Dirac} the properties of the coefficient functions lead to wave equations obeyed by the translation-matrix-invariant covariant fields. In Section \ref{Maxwell} the position dependent currents are shown to be the vector potentials of the position independent currents, for currents determined by the coefficients limited to one momentum and one helicity. Appendix A treats the adjoint representations. Appendix B contains a problem set.

\section{Wigner Little Group Representation } \label{Poincare}

The special vector for the massless particle class of Poincar\'{e} transformations is a light-like vector along $z,$ 
\begin{equation} \label{k}
k^{\mu} = \{0,0,k,k\} \quad ,
\end{equation}
with $k >$ 0, and where the first three are space components and the fourth is the time component. For convenience the units of $k$ are those of a momentum, so that the product with a displacement $x,$ $kx,$ is unitless. 

The little group $W$ of Lorentz transformations (these do not include translations) that preserves the special 4-vector $k^{\mu}$ is a combination of rotations and boosts. The generators $\{L_{1},L_{2},J\}$ of $W$ for spin 1/2 can be taken to be \cite{Weinberg,T,K2}
\begin{equation} \label{L1}
L_{1} = J^{31} + J^{14} = \frac{i}{2}\pmatrix{\sigma^{1} + i \sigma^{2} && 0 \cr 0 && -(\sigma^{1} - i \sigma^{2})} = \pmatrix{0&&i&&0&&0\cr 0&&0&&0&&0\cr 0&&0&&0&&0\cr 0&&0&&-i&&0} \quad ,
\end{equation}
\begin{equation} \label{L2}
 L_{2} = -J^{23} + J^{24} = \frac{1}{2}\pmatrix{\sigma^{1} + i \sigma^{2} && 0 \cr 0 && \sigma^{1} - i \sigma^{2}} = \pmatrix{0&&1&&0&&0\cr 0&&0&&0&&0\cr 0&&0&&0&&0\cr 0&&0&&1&&0}\quad ,
\end{equation}
and
\begin{equation} \label{J}
J = J^{12} = \frac{-1}{2}\pmatrix{\sigma^{3} && 0 \cr 0 && \sigma^{3}}= -\frac{1}{2}\pmatrix{1&&0&&0&&0\cr 0&&-1&&0&&0\cr 0&&0&&1&&0\cr 0&&0&&0&&-1} \quad ,
\end{equation}
where the matrices displayed are for the $\gamma$s in I(1). The generators $\{L_{1},L_{2},J\}$ obey the same commutation rules as the generators $\{P^x,P^y,J^z\}$ of translations and rotations in the $xy$-plane, which make up the subgroup conventionally designated $E_2.$ Of course, since $L_1$ and $L_2$ are combinations of angular momentum and boost generators, the momentum matrices $P^x$ and $P^y$ are not involved. In some unspecified, `abstract' two dimensional space $L_1$ and $L_2$ generate translations and $J$ generates rotations. 

The most general little group transformation can be written in the form 
\begin{equation} \label{W}
W(\alpha,\beta,\theta) = e^{-i(\alpha L_1 + \beta L_2)} e^{i \theta J} \quad ,
\end{equation}
which is a rotation through angle $\theta$ followed by a translation through displacement $\{\alpha,\beta\}$ in the abstract two dimensional space.

\section{Invariant Coefficient Hypothesis } \label{derive}

The calculation presented in this section closely follows paper I, which followed Weinberg, \cite{Weinberg}. 

With invariant coefficient functions, the construction $\psi$ = $\sum ua$ transforms with a \linebreak Poincar\'{e} transformation $(\Lambda , b)$ according to  
\begin{equation} \label{Basis}
\psi = \sum ua \quad \rightarrow \quad \psi^{\prime} = \sum u a^{\prime} \quad ,
\end{equation}
where a prime indicates the transformed quantity, $\psi$ is the covariant vector field, $a$ is one of the canonical vector fields and $u$ indicates the coefficient function. All the results below are based on the constraints imposed by the Invariant Coefficient Hypothesis and the transformation characters of the covariant field vector $\psi$ and the canonical field vectors $a.$

The covariant vector fields $\psi^{(12)}$ and $\psi^{(21)},$  are required to be  linear combinations of canonical field vectors $a,$  
\begin{equation} \label{psi0}
\psi^{(12)}_{l}(x) = \sum_{\sigma} \int d^3 p \enspace u^{(12)}_{l}(x;{\mathbf{p}},\sigma) a({\mathbf{p}},\sigma)  \quad 
\end{equation}
and
\begin{equation} \label{psi021}
\psi^{(21)}_{l}(x) = \sum_{\sigma} \int d^3 p \enspace u^{(21)}_{l}(x;{\mathbf{p}},\sigma) a({\mathbf{p}},\sigma)  \quad .
\end{equation}
The symbol $\mathbf{p}$ denotes the space components of the momentum, $\{p^{x},$$p^{y},$$p^{z}\}.$ The spacial components determine the time component because the mass is zero, ${\mathbf{p}}^{2}$ = ${p^{t}}^2,$ and the energy $p^t$ is required to be positive for this class of Poincar\'{e} transformations.

The covariant fields $\psi^{(12)}$ and $\psi^{(21)}$ transform like  relativistic field operators, \cite{Weinberg,T}
\begin{equation} \label{Dpsi2}
U(\Lambda,b) \psi^{(12)}_{l}(x) {U}^{-1}(\Lambda,b) = \sum_{\bar{l}} D^{(12)-1}_{l \bar{l}}(\Lambda,b)  \psi^{(12)}_{\bar{l}}(\Lambda x + b) \quad ,
\end{equation}
and
\begin{equation} \label{Dpsi2a}
U(\Lambda,b) \psi^{(21)}_{l}(x) {U}^{-1}(\Lambda,b) = \sum_{\bar{l}} D^{(21)-1}_{l \bar{l}}(\Lambda,b)  \psi^{(21)}_{\bar{l}}(\Lambda x + b) \quad ,
\end{equation}
where $D^{(12)}(\Lambda,b)$ and $D^{(21)}(\Lambda,b)$ are the spin 1/2 covariant nonunitary matrices representing the spacetime transformation $(\Lambda,b)$ in the 12- and 21-representations of the Poincar\'{e} group discussed in Section \ref{Poincare} above. The matrices transform the components labeled by the index $l$ and there is a differential representation that transforms functions defined on the space of continuous variables $x \rightarrow$ $\Lambda x + b.$ 

The canonical field vectors $a$ transform like annihilation operators and single particle states, \cite{Weinberg,T}
\begin{equation} \label{Da}
U(\Lambda,b) a({\mathbf{p}},\sigma) {U}^{-1}(\Lambda,b) = e^{-i \Lambda p \cdot b} \sqrt{\frac{(\Lambda p)^t}{p^t}}  \sum_{\bar{\sigma}} D^{({\mathrm{Can}})}_{\sigma \bar{\sigma}}(W^{-1}(\Lambda,p))  a({\mathbf{ p_{\Lambda}}},\bar{\sigma}) \quad ,
\end{equation}
where the label `Can' stands for canonical and where
\begin{equation} \label{WLpa}
W(\Lambda, p) = L^{-1}(\Lambda p)\Lambda L(p) \quad ,
\end{equation}
with $L(p) $ a standard transformation taking the special 4-vector $k^{\mu}$ = $\{0,0,k,k\}$ to $p^{\mu},$ e.g. a boost along $z$ followed by a rotation taking the unit vector $\hat{z}$ to $\hat{{\mathbf{p}}}.$ The momenta can be parameratized by
\begin{equation} \label{WLp}
p^{\mu} = k e^{\xi} \{ \cos{\phi}\sin{\theta}, \sin{\phi}\sin{\theta},\cos{\theta},1\} \quad ,
\end{equation}
where $\theta$ and $\phi$ are angles determining the direction of ${\mathbf{p}}$ and $e^{\xi}$ is the ratio of energies $p^{t}/k^{t}.$ The momenta are
restricted by
\begin{equation} \label{p2M2}
 p_{\mu}p^{\mu} = 0 \quad .
\end{equation}
Thus $p^{\mu}$ is determined by ${\mathbf{p}}$ together with $p^{4} > $ 0. The space components of the transformed momentum $\Lambda p$ are denoted ${\mathbf{ p_{\Lambda}}}.$ The matrices $D^{({\mathrm{Can}})}$ form a {\it{can}}onical representation of the little group composed of the $W(\Lambda, p)$s. A canonical representation, as used here, must have matrices dependent on $p^{\mu}$ as the form $D^{({\mathrm{Can}})}_{\sigma \bar{\sigma}}(W^{-1}(\Lambda,p))$ suggests but there is no ad hoc requirement that the representation be unitary.  

The dependence of $u^{(12)}_{l}(x;{\mathbf{p}},\sigma)$ on coordinates $x$ and translation $b$ can be assimilated by defining $u_{\bar{l}}({\mathbf{p}},\sigma)$ in
\begin{equation} \label{Du2a}
  u^{(12)}_{l}(x;{\mathbf{p}},\sigma) = (2 \pi)^{-3/2} e^{-i p \cdot x} \sum_{\bar{l}} D^{(12)}_{l \bar{l}}(1,x)  u_{\bar{l}}({\mathbf{p}},\sigma) \quad 
\end{equation}
and
\begin{equation} \label{Du2a21}
  u^{(21)}_{l}(x;{\mathbf{p}},\sigma) = (2 \pi)^{-3/2} e^{-i p \cdot x} \sum_{\bar{l}} D^{(21)}_{l \bar{l}}(1,x)  u_{\bar{l}}({\mathbf{p}},\sigma) \quad .
\end{equation}
By (\ref{Dpsi2}), (\ref{Da}), (\ref{Du2a}) and (\ref{Du2a21}), the transformed equation $\psi^{\prime}$ =  $\sum u a^{\prime}$ reduces to
\begin{equation} \label{Du2c}
    \sum_{\bar{l}} D_{l \bar{l}}(\Lambda) u_{\bar{l}}({\mathbf{p}},\sigma) = \sqrt{\frac{ (\Lambda p)^t}{p^t}} \sum_{\bar{\sigma}} u_{l}({\mathbf{ p_{\Lambda}}},\bar{\sigma}) D^{({\mathrm{Can}})}_{\bar{\sigma} \sigma}(W(\Lambda,p))    \quad .
\end{equation}
The labels $(12)$ and $(21)$ are dropped on $u_{\bar{l}}({\mathbf{p}},\sigma)$ because equation (\ref{Du2c}) for $u_{\bar{l}}({\mathbf{p}},\sigma)$ is the same equation in both the 12- and the 21-representations. Of course if there is more than one solution to the equation, then the function $u_{\bar{l}}({\mathbf{p}},\sigma)$ for the 12-representation may differ from the function $u_{\bar{l}}({\mathbf{p}},\sigma)$ for the 21-representation. 

To determine the particle spin, suppose $p^{\mu}$ is the special 4-vector, $p^{\mu}$ = $k^{\mu}$ = $\{0,0,k,k\}$  and let $\Lambda$ be a little group transformation $W_E.$ The little group transformation has no effect the special 4-vector $k^{\mu}$, by definition, and it follows that $W_E k$ = $k$ and ${\mathbf{ p_{W_{E}}}}$ = ${\mathbf{ k}}$. Also, $W(W_E,k)$ = $L^{-1}(W_E k)W_E L(k)$ = $W_E$ because $L(k)$ = $L^{-1}(k)$ = 1. In this case (\ref{Du2c}) reads
\begin{equation} \label{Du2d}
    \sum_{\bar{l}} D_{l \bar{l}}(W_E) u_{\bar{l}}({\mathbf{k}},\sigma) =  \sum_{\bar{\sigma}} u_{l}({\mathbf{k}},\bar{\sigma}) D^{({\mathrm{Can}})}_{\bar{\sigma} \sigma}(W_E)    \quad .
\end{equation}
By I(7) and I(19) with $\Lambda$ = $W_E$ and $b$ = 0, the matrix $D_{l \bar{l}}(W_E)$ has a block diagonal form and (\ref{Du2d}) implies that
\begin{equation} \label{Du2da}
    \frac{1}{2} \sum_{\bar{m}} (-\alpha L_1 - \beta L_2 + \theta J)_{m \bar{m}} u_{\bar{m} +}({\mathbf{k}},\sigma) =  \sum_{\bar{\sigma}} u_{m +}({\mathbf{k}},\bar{\sigma}) (-\alpha \lambda_1 - \beta \lambda_2 + \theta j)_{\bar{\sigma} \sigma}    \quad ,
\end{equation}
and
\begin{equation} \label{Du2da1}
    \frac{1}{2} \sum_{\bar{m}} (-\alpha L_1 - \beta L_2 + \theta J)_{m \bar{m}} u_{\bar{m} -}({\mathbf{k}},\sigma) =  \sum_{\bar{\sigma}} u_{m -}({\mathbf{k}},\bar{\sigma}) (-\alpha \lambda_1 - \beta \lambda_2 + \theta j)_{\bar{\sigma} \sigma}    \quad ,
\end{equation}
where $\lambda_1,$ $\lambda_2,$ and $j$ are the little group generators for the canonical representation. See Table 1. By one of Schur's lemmas \cite{H1} it follows that, unless the coefficient functions vanish, the generators $L_1,$ $L_2,$ and $J$ and $\lambda_1,$ $\lambda_2,$ and $j$ are equivalent, i.e. there is a similarity transformation that takes one set to the other. 

But the little group generators $L^{(11)}_1,$ $L^{(11)}_2,$ and $J^{(11)}$ of the 11-block are not equivalent to the 22-block generators $L^{(22)}_1,$ $L^{(22)}_2,$ and $J^{(22)}.$ Clearly both cannot be equivalent to the canonical generators $\lambda_1,$ $\lambda_2,$ and $j$ of which there is just one. Thus it could be that the 11-block generators are equivalent to the canonical generators and then the coefficients $u_{\bar{m} -}({\mathbf{k}},\sigma)$ are null. Therefore one possibility is that
\begin{equation} \label{jGENI}
    \{\lambda_{1}^{(I)},\lambda_{2}^{(I)},j^{(I)} \} = \{L^{(11)}_1,L^{(11)}_2,J^{(11)}\}  \quad ,
\end{equation}
which entails, by Schur's Lemma, that the lower block vanishes for the $\gamma$s in I(1),  
\begin{equation} \label{uI}
    u_{l}^{(I)}({\mathbf{k}},\sigma) = \pmatrix{u_{m+}({\mathbf{k}},\sigma)\cr 0} \quad .
\end{equation}
Or it could be that $\lambda_1,$ $\lambda_2,$ and $j$ are equivalent to the 22-block generators and the coefficients $u_{\bar{m} +}({\mathbf{k}},\sigma)$ are zero. So another possibility is that
\begin{equation} \label{jGENII}
    \{\lambda_{1}^{(II)},\lambda_{2}^{(II)},j^{(II)} \} = \{L^{(22)}_1,L^{(22)}_2,J^{(22)}\}  \quad 
\end{equation}
and the upper block vanishes,
\begin{equation} \label{uII}
    u_{l}^{(II)}({\mathbf{k}},\sigma) = \pmatrix{0 \cr u_{m-}({\mathbf{k}},\sigma)} \quad .
\end{equation}
Because either the $(0,1/2)$ right-handed (+) part of the coefficient vanishes or the $(1/2,0)$ left-handed $(-)$ part of the coefficient vanishes these coefficients have definite parity.

Knowing the generators $\lambda_1,$ $\lambda_2,$ and $j$ determines the representation $D^{(1/2)}$ which can now be used to determine $u({\mathbf{k}},\sigma).$ There are two cases distinguished by the labels $(I)$ and $(II).$ Replacing $ \{\lambda_{1}^{(I)},\lambda_{2}^{(I)},j^{(I)} \} $ in (\ref{Du2da}) with $  \{L^{(11)}_1,L^{(11)}_2,J^{(11)}\}$ implies that the coefficients $u_{m +}({\mathbf{k}},\sigma)$ form a matrix that commute with the generators $  \{L^{(11)}_1,L^{(11)}_2,J^{(11)}\}.$ One can show that $u_{m +}^{(I)}({\mathbf{k}},\sigma)$ must therefore be proportional to the unit matrix, $u_{m +}^{(I)}({\mathbf{k}},\sigma)$ = $c_{+}^{(I)} \delta_{m \sigma},$
\begin{equation} \label{Du2f}
  u_{l}^{(I)}({\mathbf{k}},+1/2) =  \pmatrix{u_{m +}^{(I)}({\mathbf{k}},+1/2) \cr 0}  = \pmatrix{c_{+}^{(I)} \delta_{m, +1/2} \cr 0} = \pmatrix{c_{+}^{(I)} \cr 0 \cr 0 \cr 0}    \quad 
\end{equation}
and 
\begin{equation} \label{Du2f1}
  u_{l}^{(I)}({\mathbf{k}},-1/2) =  \pmatrix{u_{m +}^{(I)}({\mathbf{k}},-1/2) \cr 0}  = \pmatrix{c_{+}^{(I)} \delta_{m, -1/2} \cr 0} =  \pmatrix{0 \cr c_{+}^{(I)} \cr 0 \cr 0 }   \quad ,
\end{equation}
where $c_{+}^{(I)}$ is a constant. Thus the coefficient functions for $p^{\mu}$ = $k^{\mu}$ and generators for $(I)$ are determined by the parameter $c_{+}^{(I)},$ which may be different in the 12- and 21-representations. The constant is a normalization constant.

Likewise for choice $(II)$, (\ref{jGENII}), one finds that
\begin{equation} \label{Du2fII}
  u_{l}^{(II)}({\mathbf{k}},+1/2) =  \pmatrix{0 \cr u_{m -}^{(II)}({\mathbf{k}},+1/2) }  = \pmatrix{0 \cr c_{-}^{(II)} \delta_{m, +1/2} } = \pmatrix{0 \cr 0 \cr c_{-}^{(II)} \cr 0 }    \quad 
\end{equation}
and 
\begin{equation} \label{Du2f1II}
  u_{l}^{(II)}({\mathbf{k}},-1/2) =  \pmatrix{0 \cr u_{m -}^{(II)}({\mathbf{k}},-1/2) }  = \pmatrix{0 \cr c_{-}^{(II)} \delta_{m, -1/2} } =  \pmatrix{0 \cr 0 \cr 0 \cr c_{-}^{(II)} }   \quad .
\end{equation}
Thus the coefficient functions for $p^{\mu}$ = $k^{\mu}$ and generators for  $(II)$ are determined by the parameter $c_{-}^{(II)},$ which may be different in the 12- and 21-representations.

The little group generators $L^{(11)}_1,$ $L^{(11)}_2,$ and $J^{(11)}$ of the 11-block are not equivalent to the 22-block generators $L^{(22)}_1,$ $L^{(22)}_2,$ and $J^{(22)}.$  It is however possible to make the 11- and the 22-block generators equivalent by reducing the supply of available canonical vectors $a({\mathbf{p}},\sigma)$ to a special subset labeled $a^{(III)}({\mathbf{p}},\sigma).$ To determine which canonical vectors to keep, note that for the $\gamma$s in (1) the matrices $L_{1}$ and $L_{2},$ (\ref{L1}) and (\ref{L2}), have columns 1 and 4 filled with zeros.
Therefore, define 
\begin{equation} \label{uIIIa}
  u_{l}^{(III)}({\mathbf{k}},\sigma) =  \pmatrix{ u_{m +}^{(III)}({\mathbf{k}},\sigma) \cr u_{m -}^{(III)}({\mathbf{k}},\sigma)}  = \pmatrix{c_{+,\sigma}^{(III)} \cr 0 \cr 0 \cr c_{-,\sigma}^{(III)} }    \quad ,
\end{equation}
so that the generators $L_{1}$ and $L_{2}$ yield zero,  
\begin{equation} \label{uIIIb}
  L_{1} u^{(III)}({\mathbf{k}},\sigma) =  L_{2} u^{(III)}({\mathbf{k}},\sigma) = 0    \quad ,
\end{equation}
where `$0$' stands for a column of zeros. It also follows that  
\begin{equation} \label{uIIIc}
  [e^{(-i \alpha L_{1} + \beta L_{2})}]_{l \bar{l}} u_{\bar{l}}^{(III)}({\mathbf{k}},\sigma) =  \delta_{l \bar{l}}u_{\bar{l}}^{(III)}({\mathbf{k}},\sigma) = u_{l}^{(III)}({\mathbf{k}},\sigma)    \quad .
\end{equation}
Any transformation $\exp{(-i \alpha L_{1} + \beta L_{2})}$ has the effect of multiplying by the unit matrix, so the representation is not faithful. 

Continuing with finding the subset canonical vectors  $a^{(III)}({\mathbf{p}},\sigma),$ note that equations (\ref{Du2da}) and (\ref{Du2da1}) with null $\theta$ imply that   $D^{({\mathrm{Can}})}(W^{-1}(\Lambda,p))$ is not faithful either and the canonical transformation generators vanish, $\lambda_{1}^{(III)}$ = $\lambda_{2}^{(III)}$ = 0. Thus the appropriate subset of the canonical vectors contains vectors $a^{(III)}({\mathbf{p}},\sigma)$ that are eigenvectors of $\lambda_{1}$ and $\lambda_{2}$ with eigenvalue zero. 
\begin{equation} \label{uIIIa1}
   \lambda_{1} a^{(III)}({\mathbf{p}},\sigma) = 0 \quad {\mathrm{and}} \quad  \lambda_{2} a^{(III)}({\mathbf{p}},\sigma) = 0     \quad .
\end{equation}

By expanding the reduced set of canonical vectors $a^{(III)}({\mathbf{p}},\sigma)$ over the eigenvectors of the remaining generator $j,$ so that $j a^{(III)}({\mathbf{p}},\sigma)$ = $\sigma a^{(III)}({\mathbf{p}},\sigma),$ one can simplify the transformation rule, (\ref{Da}), for the canonical vectors, \cite{Weinberg+}
\begin{equation} \label{DaS}
U^{(III)}(\Lambda,b) a^{(III)}({\mathbf{p}},\sigma) {{U}^{(III)}}^{-1}(\Lambda,b) = e^{-i \Lambda p \cdot b} \sqrt{\frac{(\Lambda p)^t}{p^t}}  e^{-i \sigma \theta(\Lambda,p)}  a^{(III)}({\mathbf{ p_{\Lambda}}},\sigma) \quad .
\end{equation}
And equations (\ref{Du2da}) and (\ref{Du2da1}) become 
\begin{equation} \label{Du2daIII}
    \frac{1}{2} \sum_{\bar{m}} \sigma^{3}_{m \bar{m}} u_{\bar{m} +}^{(III)}({\mathbf{k}},\sigma) = \frac{1}{2} \pmatrix{1&&0 \cr 0 && -1} \pmatrix{c_{+,\sigma}^{(III)} \cr 0} =  \sigma u_{m +}^{(III)}({\mathbf{k}},\sigma)     \quad ,
\end{equation}
and
\begin{equation} \label{Du2da1III}
     \frac{1}{2} \sum_{\bar{m}} \sigma^{3}_{m \bar{m}} u_{\bar{m} -}^{(III)}({\mathbf{k}},\sigma) = \frac{1}{2} \pmatrix{1&&0 \cr 0 && -1} \pmatrix{0 \cr c_{-,\sigma}^{(III)} } =  \sigma u_{m -}^{(III)}({\mathbf{k}},\sigma)     \quad .
\end{equation}
Equation (\ref{Du2daIII}) reduces to $\sigma$ = $+1/2$ and equation (\ref{Du2da1III}) reduces to $\sigma$ = $-1/2.$ Thus $u_{l}^{(III)}({\mathbf{k}},\sigma)$ must be given by
\begin{equation} \label{uIIId}
  u_{l}^{(III)}({\mathbf{k}},+1/2) =  \pmatrix{ u_{m +}^{(III)}({\mathbf{k}},+1/2) \cr u_{m -}^{(III)}({\mathbf{k}},+1/2)}  = \pmatrix{c_{+}^{(III)} \cr 0 \cr 0 \cr 0 }    \quad 
\end{equation}
and
\begin{equation} \label{uIIIe}
  u_{l}^{(III)}({\mathbf{k}},-1/2) =  \pmatrix{ u_{m +}^{(III)}({\mathbf{k}},-1/2) \cr u_{m -}^{(III)}({\mathbf{k}},-1/2)}  = \pmatrix{0 \cr 0 \cr 0 \cr c_{-}^{(III)} }    \quad ,
\end{equation}
where $c_{+}^{(III)}$ = $c_{+,+1/2}^{(III)}$ and $c_{-}^{(III)}$ = $c_{-,-1/2}^{(III)}$ are the nonzero constants. Of course, the constants $c$ in  (\ref{uIIId}) and (\ref{uIIIe}) may be different for the 12- and 21-representations.

In summary, equations (\ref{Du2da}) and (\ref{Du2da1}) have lead to specifying three different canonical transformation rules labeled by $I,II,III.$ Let the index $A$ be used to indicate one of these representations, $A \in$ $\{I,II,III\}.$ Each canonical transformation rule has its own set of generators $\{ \lambda_1,\lambda_2,j \}^{(A)}.$  Each has its own set of coefficients $u^{(A)}({\mathbf{k}},\sigma)$ and, since the canonical vectors transform differently for $I, II, III,$ there are three sets of canonical vectors, $a^{(A)}({\mathbf{p}},\sigma).$

To find the $u({\mathbf{p}},\sigma)$ in (\ref{Du2a}) consider (\ref{Du2c}) when $\Lambda$ = $L^{-1}(p).$ Since $L(p)$ takes $k$ to $p,$ it follows that $L^{-1}(p)$ takes $p$ to $k$ and that $W(L^{-1}(p),p)$ = $L^{-1}(L^{-1}(p)p)L^{-1}(p)L(p)$ = $L^{-1}(k)$ = 1. Now (\ref{Du2c}) becomes, for this case, 
\begin{equation} \label{Du2g}
     u_{l}^{(A)}({\mathbf{p}},\sigma) = \sqrt{\frac{ k}{p^t}} \sum_{\bar{l}}D_{l \bar{l}}(L(p))  u_{\bar{l}}^{(A)}({\mathbf{k}},\sigma)     \quad .
\end{equation}

By (\ref{Du2a}), (\ref{Du2a21}), (\ref{Du2f}), and (\ref{Du2g}) the coefficient functions $u^{(12;A)}_{l}(x;{\mathbf{p}},\sigma)$ and $u^{(21;A)}_{l}(x;{\mathbf{p}},\sigma)$ are given by 
\begin{equation} \label{Du2h}
  u^{(12;A)}_{l}(x;{\mathbf{p}},\sigma) = D^{(12)}_{l \bar{l}}(1,x) u_{\bar{l}}^{(A)}(x;{\mathbf{p}},\sigma) \quad \end{equation}
and
\begin{equation} \label{Du2h1}
u^{(21;A)}_{l}(x;{\mathbf{p}},\sigma) = D^{(21)}_{l \bar{l}}(1,x) u_{\bar{l}}^{(A)}(x;{\mathbf{p}},\sigma) \quad ,
\end{equation}
where $u^{(A)}(x;{\mathbf{p}},\sigma)$ is given by
\begin{equation} \label{Du2i}
  u_{\bar{l}}^{(A)}(x;{\mathbf{p}},\sigma) = (2 \pi)^{-3/2}\sqrt{\frac{ k}{p^t}} e^{-i p \cdot x} \sum_{n} D_{\bar{l}n}(L(p))  u_{n}^{(A)}({\mathbf{k}},\sigma) \quad ,
\end{equation}
for both the 12- and 21-representations.    

The structure of the expressions in (\ref{Du2h}), (\ref{Du2h1}) and (\ref{Du2i}) reflect the observation that the 12- and 21- representations agree for rotations and boosts, i.e. note the $D(L(p))$ in the expression (\ref{Du2i}) for $u^{(A)}(x;{\mathbf{p}},\sigma),$ but the representations differ for translations, i.e. note the $D^{(12)}(1,x)$ and $D^{(21)}(1,x)$ in (\ref{Du2h}) and (\ref{Du2h1}) that distinguish $u^{(12;A)}(x;{\mathbf{p}},\sigma)$ from $u^{(21;A)}(x;{\mathbf{p}},\sigma).$

%\pagebreak

\section{Relating the 12- and 21-Fields Fails} \label{1221field}

The reason for the difference in the parity behavior of the massive and massless classes of Poincar\'{e} transformations is the parity behavior of the respective little groups. The $(0,1/2)$ right-handed representation of the massive little group is equivalent to the $(1/2,0)$ left-handed representation since the little group consists of rotations only, no boosts. But the little group of the massless class combines rotations and boosts which are represented differently in the right and left-handed representations $(0,1/2)$ and $(1/2,0).$

As discussed in Section I3, the 12- and 21-representations of the Poincar\'{e} group of spacetime transformations are related by the 12/21 transition, i.e. a similarity transformation and an exchange of contravariant and covariant indices as displayed in I(34). As discussed there, this amounts to a parity transformation. For massive particles, this relationship induces a  relationship between 12- and 21-fields. For massless particles, as will be shown in this section, no relationship between the coefficient functions for any one transformation character $A$ is induced by the 12/21 transition. In fact assuming that such a relationship exists between a 12 coefficient function $u^{(12)}$ and a 21-coefficient function $u^{(21)}$ implies that both coefficient functions vanish.

If the 12/21 transition induces a relationship between coefficients $u^{(12;A)}$ and $\tilde{u}^{(21;A)}$, then, by I(34), I(35), (\ref{Du2h}) and (\ref{Du2i}), one finds just as in paper I that
\begin{equation} \label{1221a}
\gamma^{4}  u^{(12;A)}(x,{\mathbf{p}},\sigma) = \tilde{u}^{(21;A)}(\tilde{x},\tilde{{\mathbf{p}}},\sigma) \quad , \hspace{5.5cm}
\end{equation}
where
\begin{equation} \label{1221b}
 \tilde{u}^{(21;A)}(\tilde{x},\tilde{{\mathbf{p}}},\sigma)= (2 \pi)^{-3/2}\sqrt{\frac{ k}{\tilde{p}^t}}  e^{-i \tilde{p} \cdot \tilde{x}} D^{(21)}(L(\tilde{p}),\tilde{x})  \tilde{u}^{(A)}({\mathbf{k}},\sigma)\quad .\end{equation}
In these equations,
\begin{equation} \label{1221c}
 \tilde{p}^{\mu} = \eta_{\mu \nu} p^{\nu} = p_{\mu} \quad {\mathrm{and}} \quad \tilde{x}^{\mu} =  x_{\mu} \quad 
\end{equation}
and
 \begin{equation} \label{1221d}
 \tilde{u}({\mathbf{k}},\sigma) = \gamma^{4} u({\mathbf{k}},\sigma)  \quad . \end{equation}

With $A$ = $I,$ $\sigma$ = $+1/2$ and the $\gamma$s in I(1), equation (\ref{1221d}) becomes
  \begin{equation} \label{1221d1}
 \tilde{u}^{(I)}({\mathbf{k}},1/2) = \pmatrix{\tilde{c}_{+}^{(I)} \cr 0 \cr 0 \cr 0} = \gamma^{4} u^{(I)}({\mathbf{k}},1/2) = \pmatrix{0 \cr 0 \cr c_{+}^{(I)} \cr 0} \quad  \end{equation}
and, for $\sigma$ = $-1/2$,
  \begin{equation} \label{1221d1b}
 \tilde{u}^{(I)}({\mathbf{k}},-1/2) = \pmatrix{0 \cr \tilde{c}_{+}^{(I)} \cr 0 \cr 0 } = \gamma^{4} u^{(I)}({\mathbf{k}},-1/2) = \pmatrix{0 \cr 0 \cr 0 \cr c_{+}^{(I)}} \quad . \end{equation}
Thus the parameter $c_{+}^{(I)}$ for $u^{(I)}({\mathbf{k}},\sigma)$ and the parameter  $\tilde{c}_{+}^{(I)}$ for $\tilde{u}^{(I)}({\mathbf{k}},\sigma),$ see (\ref{Du2f}) and (\ref{Du2f1}), must vanish 
\begin{equation} \label{1221e}
\tilde{c}_{+}^{(I)} = c_{+}^{(I)} = 0. \quad 
\end{equation}
By (\ref{Du2h}), (\ref{Du2h1}) and (\ref{Du2i}), the coefficients must vanish as well,
\begin{equation} \label{1221e1}
u^{(12;A)}(x,{\mathbf{p}},\sigma) = \tilde{u}^{(21;A)}(\tilde{x},\tilde{{\mathbf{p}}},\sigma) = 0 \quad .
\end{equation}
Thus there is no nonzero coefficient function $\tilde{u}^{(21;I)}(\tilde{x},\tilde{{\mathbf{p}}},\sigma)$ corresponding to any coefficient function $u^{(12;I)}(x,{\mathbf{p}},\sigma)$ by way of the 12/21 transition. 

The same result follows for the $II$ and $III$ fields; no nonzero coefficient function \linebreak $u^{(12;A)}(x,{\mathbf{p}},\sigma)$ has a related coefficient function $\tilde{u}^{(21;A)}(\tilde{x},\tilde{{\mathbf{p}}},\sigma).$  Therefore the relationship between the 12- and 21-representations discussed in Section I3 induces no relationship between 21-fields $\tilde{\psi}^{(21;A)}$ and 12-fields $\psi^{(12;A)}.$ 

For $\gamma$s in the form I(1), the 12/21 transition similarity matrix $\gamma^{4}$ = $\pmatrix{0 && 1 \cr 1 && 0}$ exchanges the upper block of two components of a coefficient with the lower block of two components. But for $I$ and $II,$  Schur's Lemma makes one block vanish because the 11- and 22-representations of the massless little group are not equivalent. Thus the 12/21 transition similarity matrix $\gamma^{4}$ equates a nonzero block to a zero block making for a null result. Thus the similarity matrix $\gamma^{4}$ exchanges parity and, for $I$ and $II,$ Schur's Lemma forces fixed parity. Requiring both yields null results.

For $III$ neither block is zero, but the result is null anyway because the spins of the right-hand block $u^{(III)}_{+}$ is opposite the spin for the left-hand block $u^{(III)}_{-}.$

\section{Translation Matrix Invariance; Wave Equations} \label{Dirac}

The covariant translation transformations $\psi^{(12)}_{l}(x) \rightarrow$ $D^{(12)-1}_{l \bar{l}}(1,b)\psi_{\bar{l}}(x+b),$ and $\psi^{(21)}_{l}(x) \rightarrow$ $D^{(21)-1}_{l \bar{l}}(1,b)\psi_{\bar{l}}(x+b),$ I(40) and I(41) with no rotations or boosts, represent a translation through displacement $b$ both as a finite dimensional matrices $D^{(12)}(1,b)$ and $D^{(21)}(1,b)$ as well as a differential representation that takes functions of $x$ to $x + b.$ In this section fields invariant to the application of the matrix representation of translations are discussed. Fields that are not invariant in this way are discussed in the next section.

It is easy to see from the matrices for $P_{(12)}^{\mu}$ and $P_{(21)}^{\mu}$ in I(9) and I(11) that $u^{(12;I)},$ $u^{(21;II)},$ $u^{(12;III)}(x,{\mathbf{p}},+1/2)$ and $u^{(21;III)}(x,{\mathbf{p}},-1/2)$  are invariant under matrix translations. The momentum matrices $P_{(12)}^{\mu}$  have block triangular form $\pmatrix{0 && 1 \cr 0 && 0 }$ and the $P_{(21)}^{\mu}$ have the form $\pmatrix{0 && 0 \cr 1 && 0 },$ while the coefficient functions $u^{(A)}(x;{\mathbf{p}},\sigma),$ $A\in$ $\{I,II,III\},$ have block form with either the lower block of two components zero $\pmatrix{1 \cr 0}$ or the upper block zero $\pmatrix{0 \cr 1}$. Thus the coefficients $u^{(A)}(x;{\mathbf{p}},\sigma),$ in the form $\pmatrix{1 \cr 0},$ i.e. $u^{(12;I)}(x;{\mathbf{p}},\sigma)$ and $u^{(12;III)}(x,{\mathbf{p}},+1/2),$ are invariant upon multiplication by 12-translation matrices
\begin{equation} \label{TMI1}
 \pmatrix{1 && -ibP \cr 0 && 1} \pmatrix{u_{+} \cr 0} = \pmatrix{u_{+} \cr 0} \quad .\end{equation}
And those of the form $\pmatrix{0 \cr 1},$ i.e. $u^{(21;II)}(x;{\mathbf{p}},\sigma)$ and $u^{(21;III)}(x,{\mathbf{p}},-1/2),$ are invariant upon multiplication by 21-translation matrices.

Translation matrix invariance simplifies the transformation of the corresponding covariant fields by replacing $ D^{-1}_{l \bar{l}}(\Lambda,b)$ by $D^{-1}_{l \bar{l}}(\Lambda)$ in (\ref{Dpsi2}) and (\ref{Dpsi2a}),
\begin{equation} \label{PO1}
 U(\Lambda,b) \psi^{(12;I)}_{l}(x) {U}^{-1}(\Lambda,b) =  D^{-1}_{l \bar{l}}(\Lambda)\psi^{(12;I)}_{\bar{l}}(\Lambda x + b) \quad ,\end{equation}
\begin{equation} \label{PO2}
 U(\Lambda,b) \psi^{(21;II)}_{l}(x) {U}^{-1}(\Lambda,b) =  D^{-1}_{l \bar{l}}(\Lambda)\psi^{(21;II)}_{\bar{l}}(\Lambda x + b) \quad ,\end{equation}
\begin{equation} \label{PO3a}
 U(\Lambda,b) \psi^{(12;III)}_{l}(x) {U}^{-1}(\Lambda,b)\mid_{\sigma = +1/2} =  D^{-1}_{l \bar{l}}(\Lambda)\psi^{(12;III)}_{\bar{l}}(\Lambda x + b)\mid_{\sigma = +1/2}  \quad ,\end{equation}
\begin{equation} \label{PO3b}
 U(\Lambda,b) \psi^{(21;III)}_{l}(x) {U}^{-1}(\Lambda,b)\mid_{\sigma = -1/2} =  D^{-1}_{l \bar{l}}(\Lambda)\psi^{(21;III)}_{\bar{l}}(\Lambda x + b)\mid_{\sigma = -1/2}  \quad .\end{equation}
Thus, for these fields, the components are not rearranged upon translation; the dependence on coordinates $x^{\mu}$ occurs just in the form of phase factors $\exp{(-i p \cdot x)}$ in the coefficients $u^{(A)}$ as displayed in (\ref{Du2i}). Coordinate dependence just in phase factors, i.e. plane waves, enables the following method of deriving wave equations.  

To begin with, suppose there is a matrix $M$ that produces zero when acting on one of the translation matrix invariant fields, say, $u^{(12;I)}({\mathbf{k}},\sigma),$ 
\begin{equation} \label{Wave1}
 M_{l \bar{l}} u_{\bar{l}}^{(12;I)}({\mathbf{k}},\sigma) = 0 \quad . \end{equation}
Then, by (\ref{Du2h}), (\ref{Du2h1}), (\ref{Du2i}) and since $u^{(12;I)}$ is invariant under matrix translations, one finds that
\begin{equation} \label{Wave2}
 D(L(p))M_{l \bar{l}} D^{-1}(L(p)) u_{\bar{l}}^{(12;I)}(x;{\mathbf{p}},\sigma) = 0 \quad ,\end{equation}
which is a start on a wave equation.

But $\psi^{(12;I)}(x)$ is constructed from a linear combination of terms proportional to \linebreak $u^{(12;I)}(x;{\mathbf{p}},\sigma).$ In order to deduce a linear differential equation from the relation (\ref{Wave2}) note that each coefficient $u^{(12;I)}(x;{\mathbf{p}},\sigma)$ depends on the coordinates $x^{\mu}$ in the phase factor \linebreak $\exp{(-i p_{\mu} x^{\mu})}$ and in no other way. Thus the operator $i \partial_{\mu}$ brings down a factor of $p_{\mu}$ when applied to any of the coefficients.  A common linear differential expression, call it $\Pi(i \partial_{\mu})$ can be applied to $\psi^{(12;I)}(x)$ and it will act coefficient-by-coefficient. Thus in order to ensure that (\ref{Wave2}) leads to a differential equation for $\psi^{(12;I)}(x),$ it suffices to require that
\begin{equation} \label{Wave3}
 D(L(p))M D^{-1}(L(p))  = \Pi(p_{\mu}) \quad .\end{equation}
Note that this is not a necessary requirement. 

If one can find a matrix-valued function $\Pi(p_{\mu}),$ it follows that 
\begin{equation} \label{Wave4}
 \Pi(i \partial_{\mu}) \psi^{(12;I)}(x) = 0 \quad ,\end{equation}
because applying the linear differential expression $\Pi(i \partial_{\mu})$ to $\psi^{(12;I)}(x)$ multiplies each coefficient $u_{\bar{l}}^{(12;I)}(x;{\mathbf{p}},\sigma)$ by the matrix $\Pi(p_{\mu})$ and by (\ref{Wave2}) the result vanishes. Equation (\ref{Wave4}) has the form of a wave equation.

To find $\Pi(p_{\mu}),$ note that $L(p)$ takes the standard 4-vector $k$ to $p,$ i.e. $L(p)k$ = $p$ with  $L(k)$ = 1, and it follows from (\ref{Wave3}) that $M$ = $\Pi(k).$  Thus (\ref{Wave3}) can be rewritten as
\begin{equation} \label{mp1}
 D(L(p))\Pi(k) D^{-1}(L(p))  = \Pi(p) \quad \end{equation}
and, by (\ref{Wave2}), one finds that
\begin{equation} \label{mp1a}
 D(L(p))\Pi(k) D^{-1}(L(p))u^{(12;I)}(x;{\mathbf{p}},\sigma)    = \Pi(p)u^{(12;I)}(x;{\mathbf{p}},\sigma) = 0  \quad .\end{equation}

The defining property of vector matrices is that vector matrices are both 4-vectors and second order tensors, i.e.
\begin{equation} \label{vecTEN}
D^{-1}(\Lambda)\gamma^{\mu} D(\Lambda) = \Lambda^{\mu}_{\nu} \gamma^{\nu}.
\end{equation}
For $\Lambda$ = $L^{-1}(p),$ this reads 
\begin{equation} \label{vecTEN1}
D(L(p))\gamma^{\mu} D^{-1}(L(p)) = {L^{-1}}^{\mu}_{\nu}(p) \gamma^{\nu}.
\end{equation}
Comparing (\ref{mp1a}) and (\ref{vecTEN1}) shows the same structure in both equations.

Assume that $\Pi(p)$ is a scalar product of the $\gamma$s with some vector function of $p^{\mu},$
\begin{equation} \label{mp1b}
  \Pi(p) = \pi_{\mu}(p)\gamma^{\mu}  \quad ,\end{equation}
where $\pi_{\mu}(p) $ is an as-yet-unknown 4-vector-valued function.
This assumption does not give the most general solutions to (\ref{mp1a}), but it does give first order equations and it has the advantage of simplicity.

The problem can now be related to the properties of currents. Upon multiplying (\ref{mp1a}) by $u^{\dagger} \gamma^{4},$ and since $D^{-1}(\Lambda)$ = $D(\Lambda^{-1}),$  $D^{\dagger}(\Lambda)$ = $\gamma^{4}D(\Lambda)\gamma^{4},$ the form (\ref{mp1b}) implies that
\begin{equation} \label{mp1c}
  \pi_{\mu}(k)u^{(12;I;\dagger)}(x;{\mathbf{p}},\sigma) {D^{-1}}^{\dagger}(L^{-1}(p))\gamma^{4} \gamma^{\mu} D(L^{-1}(p))u^{(12;I)}(x;{\mathbf{p}},\sigma) = 0 \quad .\end{equation}
By the relation giving the coefficient for $p^{\mu}$ from the coefficient for $k^{\mu},$ (\ref{Du2i}), this simplifies somewhat to
 \begin{equation} \label{mp1d}
  \pi_{\mu}(k)u^{(12;I;\dagger)}(0;{\mathbf{k}},\sigma) \gamma^{4} \gamma^{\mu} u^{(12;I)}(0;{\mathbf{k}},\sigma) = 0 \quad . \end{equation}
This form can be further simplified by introducing `currents'.

Define the currents $j^{(12;A) \mu}$ and $j^{(21;A) \mu}$ 
\begin{equation} \label{CVP1}
  j^{(12;A) \mu}(x;{\mathbf{p}},\sigma) = p^{t}\bar{u}^{(12;A)}(x;{\mathbf{p}},\sigma)\gamma^{\mu} u^{(12;A)}(x;{\mathbf{p}},\sigma) \quad 
\end{equation}
and
\begin{equation} \label{CVP2}
  j^{(21;A) \mu}(x;{\mathbf{p}},\sigma) = p^{t}\bar{u}^{(21;A)}(x;{\mathbf{p}},\sigma)\gamma^{\mu} u^{(21;A)}(x;{\mathbf{p}},\sigma) \quad ,
\end{equation}
where $\bar{u}$ = $u^{\dagger} \gamma^{4},$ the factor $p^{t}$ adjusts the normalization so that the currents are four-vectors and $A \in$ $\{I,II,III\}$ distinguishes the various transformation characters of the canonical vectors $a^{(A)}.$

The plane wave coordinate dependence $\exp(-i p\cdot x)$ cancels out in  $j^{(12;A) \mu}$ and $j^{(21;A) \mu},$ so the only dependence on coordinates can come from the translation matrices $D^{(12)}(1,x)$ and $D^{(21)}(1,x)$ and their adjoints. Since some coefficient functions are translation-matrix-invariant, i.e. $u^{(12;I)},$ $u^{(21;II)},$ $u^{(12;III)}\mid_{\sigma = +1/2}$ and $u^{(21;III)}\mid_{\sigma = -1/2},$ it follows that these currents are constant in space and time, i.e. $j^{(12;I) \mu},$ $j^{(21;II) \mu},$ $j^{(12;III) \mu}\mid_{\sigma = +1/2}$ and $j^{(21;III) \mu}\mid_{\sigma = -1/2}$ are constant over space and time. For example, by (\ref{Du2f}), (\ref{Du2f1}), (\ref{Du2h}), and (\ref{Du2i}), the currents for $u^{(12;I)}$ are given by
\begin{equation} \label{j12I}
  j^{(12;I) \mu}(x;{\mathbf{p}},+1/2) = \frac{{c_{+}^{(I)}}^2}{(2 \pi)^3}\enspace p^{\mu}   \quad 
\end{equation}
and
\begin{equation} \label{j12Im}
  j^{(12;I) \mu}(x;{\mathbf{p}},-1/2) = \frac{{c_{+}^{(I)}}^2}{(2 \pi)^3}\frac{k^2}{{p^t}^2}\enspace p_{\mu}   \quad ,
\end{equation}
with similar expressions for the other constant currents. It is $j^{(12;I)}\mid_{\sigma = -1/2}$ and $j^{(21;II)}\mid_{\sigma = +1/2}$ that have the more complicated form (\ref{j12Im}) while the others have the simpler form (\ref{j12I}). 

One can show that the constant currents for $p^{\mu}$ are related to the constant currents for $k^{\mu},$ for example,
\begin{equation} \label{j12Ia}
  j^{(12;I) \mu}(x;{\mathbf{p}},+1/2) = L^{\mu}_{\nu}(p)j^{(12;I) \nu}(x;{\mathbf{k}},+1/2) \quad ,
\end{equation}
where $L^{\mu}_{\nu}(p)$ is the standard transformation that takes the special 4-vector $k^{\nu}$ = $\{0,0,k,k\}$ to $p^{\mu}.$ From (\ref{j12Ia}) and the same relationship for $j^{(21;I) \mu},$ it follows that $j^{(12;I) \mu}$ and $j^{(21;I) \mu}$ transforms as 4-vectors and, since this is the massless class with $p_{\mu}p^{\mu}$ = 0, they are light-like 4-vectors.
 
By the definition of the currents, (\ref{CVP1}) and (\ref{CVP2}), equation (\ref{mp1d}) can be rewritten as follows,
\begin{equation} \label{mp1dd}
  \pi_{\mu}(k) j^{(12;I) \mu}(0;{\mathbf{k}},\sigma) = 0 \quad .\end{equation}
Since the constant currents are light-like 4-vectors, one solution for $\pi_{\mu}(p)$ is just the current itself,
\begin{equation} \label{mp1e}
  \pi_{\mu}(p) = j^{(12;I)}_{\mu}(x;{\mathbf{p}},\sigma)  \quad ,\end{equation}
where the $x$-dependence is an illusion because the phase factors cancel in the current and the coefficients are translation-matrix-invariant.

Thus there are two solutions depending on the form of the current, either (\ref{j12I}) or (\ref{j12Im}),
\begin{equation} \label{mp1f}
  \pi_{\mu}(p) = \frac{{c_{+}^{(I)}}^2}{(2 \pi)^3}\enspace p_{\mu}  \quad \end{equation}
or
\begin{equation} \label{mp1g}
  \pi_{\mu}(p) = \frac{{c_{+}^{(I)}}^2}{(2 \pi)^3}\frac{k^2}{{p^t}^2}\enspace \eta^{\mu \nu} p_{\nu}  \quad ,\end{equation}
where the indices are raised or lowered as required in going from $j^{\mu}$ to $j_{\mu}.$ Also both are written in terms of the lower index $p_{\mu}$ to ease the replacement $p_{\mu} \rightarrow$ $i \partial_{\mu}.$ As discussed above the partial brings down the momentum component when it acts on the phase $\exp{(-i x \cdot p)}.$

The form (\ref{mp1f}) works for  $u^{(I)}({\mathbf{k}},+1/2),$ $u^{(II)}({\mathbf{k}},-1/2),$ $u^{(III)}({\mathbf{k}},+1/2)$ and $u^{(III)}({\mathbf{k}},-1/2),$ and gives the following wave equations,
\begin{equation} \label{WE7}
 i \gamma^{\mu} \partial_{\mu} \psi^{(12;I)}(x) \mid_{\sigma = +1/2} = 0  \quad ,\end{equation}
\begin{equation} \label{WE8}
 i \gamma^{\mu} \partial_{\mu} \psi^{(21;II)}(x) \mid_{\sigma = -1/2} = 0  \quad ,\end{equation}
\begin{equation} \label{WE9}
 i \gamma^{\mu} \partial_{\mu} \psi^{(12;III)}(x) \mid_{\sigma = +1/2} = 0  \quad ,\end{equation}
\begin{equation} \label{WE10}
 i \gamma^{\mu} \partial_{\mu} \psi^{(21;III)}(x) \mid_{\sigma = -1/2} = 0  \quad ,\end{equation}
where the initial factors in (\ref{mp1f}) are canceled because the right-hand-side is zero. These equations are clearly covariant and because the fields can have but two nonzero components, these equations are equivalent to Weyl equations.\cite{Weylequations}

The form (\ref{mp1g}) works for  $u^{(I)}({\mathbf{k}},-1/2)$ and $u^{(II)}({\mathbf{k}},+1/2),$ yielding the following wave equations,
\begin{equation} \label{Wave4NC}
 i \delta^{\mu \nu}  \gamma_{\mu} \partial_{\nu} \psi^{(12;I)}(x) \mid_{\sigma = -1/2} = 0  \quad ,\end{equation}
\begin{equation} \label{Wave5NC}
 i \delta^{\mu \nu}  \gamma_{\mu} \partial_{\nu} \psi^{(21;II)}(x) \mid_{\sigma = +1/2} =0  \quad ,\end{equation}
where the $\delta$ function arises because $\delta^{\mu \nu}\gamma_{\mu} \partial_{\nu}$ = $\eta^{\mu \alpha} \eta^{\mu \beta} \gamma_{\beta} \partial_{\alpha} $ = $\eta^{\mu \alpha} p_{\alpha} \gamma^{\mu}.$  These are not covariant; when the factor ${k^{t}}^2/{p^{t}}^2$ in (\ref{mp1g}) was canceled in (\ref{mp1dd}), the covariance was lost. See the problem set in Appendix B for other forms of these equations. 
 
In many contexts, the term `wave equation' includes the requirement of covariance.\cite{TUNG,TUNG2} Since (\ref{Wave4NC}) and (\ref{Wave5NC}) are not covariant, they would not be wave equations in that sense. However they follow from the Invariant Coefficient Hypothesis just as the more traditional wave equations do. In this paper the results follow from the Invariant Coefficient Hypothesis and the transformation rules. Covariance is not required.

The six translation-matrix-invariant fields satisfy wave equations (\ref{WE7}) - (\ref{WE10}),  (\ref{Wave4NC}) and (\ref{Wave5NC}).

%\pagebreak

\section{Current as Vector Potential; Maxwell's Equations} \label{Maxwell}

The preceding section dealt with translation-matrix-{\it{invariant}} coefficient functions. This section shows that the currents of translation-matrix-{\it{dependent}} coefficients are the vector potentials of the currents of those invariant coefficient functions. The proposition is proved by showing that the Maxwell equation for vector potential and source is satisfied by the various translation dependent and independent currents. The currents that change upon translation are the vector potentials; the sources are the currents that do not change upon translation.  

Since currents are quadratic in coefficient functions, the results do not extend immediately to fields. Since the covariant fields $\psi$ are sums over canonical fields $a$ with coefficients $u,$ and the canonical fields could be annihilation operators with special commutation properties, calculations involving fields would depend on properties of the canonical fields other than their transformation character. Such calculations are beyond the limited scope of this paper. Thus the currents are considered here on a coefficient-by-coefficient basis.

The translation-matrix-{\it{dependent}} coefficients have currents, (\ref{CVP1}) and (\ref{CVP2}), that depend on position coordinates $x^{\mu}$ due to the translation matrices $D(1,x)$ in (\ref{Du2h}) and (\ref{Du2h1}),
\begin{equation} \label{CVP3}
  j^{(12;A) \mu}(x;{\mathbf{p}},\sigma) = p^{t}{u^{(A)}}^{\dagger}(0;{\mathbf{p}},\sigma){D^{(12)}}^{\dagger}(1,x) \gamma^{4}\gamma^{\mu} D^{(12)}(1,x) {u^{(A)}}(0;{\mathbf{p}},\sigma) \quad 
\end{equation}
and
\begin{equation} \label{CVP4}
  j^{(21;A) \mu}(x;{\mathbf{p}},\sigma) = p^{t}{u^{(A)}}^{\dagger}(0;{\mathbf{p}},\sigma){D^{(21)}}^{\dagger}(1,x) \gamma^{4}\gamma^{\mu} D^{(21)}(1,x) {u^{(A)}}(0;{\mathbf{p}},\sigma) \quad .
\end{equation}
The coordinate dependence is in the following expressions
\begin{equation} \label{CVP3A}
 {D^{(12)}}^{\dagger}(1,x) \gamma^{4}\gamma^{\mu} D^{(12)}(1,x)  \quad 
\end{equation}
and
\begin{equation} \label{CVP4A}
 {D^{(21)}}^{\dagger}(1,x) \gamma^{4}\gamma^{\mu} D^{(21)}(1,x)  \quad 
\end{equation}
and these same expressions appeared with the currents in the massive case in paper I. They have the same differential properties as they had with massive particles and they lead here again to the Maxwell equations.

By a straightforward calculation, detailed in paper I, one finds that the currents obey the equations
\begin{equation} \label{CVP9}
 \partial^{\tau} \partial_{\tau}  j^{(21;I) \mu}(x;{\mathbf{p}},\sigma) - \partial^{\mu} \partial_{\kappa}j^{(21;I) \kappa}(x;{\mathbf{p}},\sigma) = -12{K}^2 \frac{{c_{+}^{(21)}}^2}{{c_{+}^{(12)}}^2}  j^{(12;I) \mu}(x;{\mathbf{p}},\sigma) \quad ,
\end{equation}
\begin{equation} \label{CVP10}
  \partial^{\tau} \partial_{\tau} j^{(12;II) \mu}(x;{\mathbf{p}},\sigma) -\partial^{\mu} \partial_{\kappa} j^{(12;II) \kappa}(x;{\mathbf{p}},\sigma)= -12{K}^2 \frac{{c_{-}^{(12)}}^2}{{c_{-}^{(21)}}^2}  j^{(21;II) \mu}(x;{\mathbf{p}},\sigma) \quad ,
\end{equation}
\begin{equation} \label{CVP11}
 \partial^{\tau} \partial_{\tau}  j^{(21;III) \mu}(x;{\mathbf{p}},+1/2) - \partial^{\mu} \partial_{\kappa}j^{(21;III) \kappa}(x;{\mathbf{p}},+1/2) = -12{K}^2 \frac{{c_{+}^{(21)}}^2}{{c_{+}^{(12)}}^2}  j^{(12;III) \mu}(x;{\mathbf{p}},+1/2) \quad ,
\end{equation}
\begin{equation} \label{CVP12}
  \partial^{\tau} \partial_{\tau} j^{(12;III) \mu}(x;{\mathbf{p}},-1/2) -\partial^{\mu} \partial_{\kappa} j^{(12;III) \kappa}(x;{\mathbf{p}},-1/2)= -12{K}^2 \frac{{c_{-}^{(21)}}^2}{{c_{-}^{(12)}}^2}  j^{(21;III) \mu}(x;{\mathbf{p}},-1/2) \quad ,
\end{equation}
where $\partial^{\tau}$ = $\eta^{\tau \nu} \partial/\partial{x^{\nu}}.$ The currents on the left sides of (\ref{CVP9})-(\ref{CVP12}) depend on $x^{\mu}$ with constant second partials. On the right sides of (\ref{CVP9})-(\ref{CVP12}) are constant currents.

Note that to make the constant on the right side, i.e. the `charge,' in (\ref{CVP9}) vanish while keeping nonzero momentum matrices $K \neq$ 0, one would need to choose $c_{+}^{(21)} $ = 0, and that would make the position dependent current $ j^{(21;I) \mu}$ vanish as well. So if there is a nonzero position dependent current, then the corresponding position independent field carries nonzero charges.

Define the vectors ${a^{(A)}}^{\mu},$ $A \in$ $\{I,II,III\},$ as proportional to the sum of the currents, 
\begin{equation} \label{CVP14}
  {a^{(I)}}^{\mu}(x;{\mathbf{p}},\sigma) = \frac{-q {c_{+}^{(12;I)}}^{2}}{12 K^2{c_{+}^{(21;I)}}^{2}} [j^{(12;I) \mu}(x;{\mathbf{p}},\sigma) + j^{(21;I) \mu}(x;{\mathbf{p}},\sigma)] \quad ,
\end{equation}
\begin{equation} \label{CVP15}
  {a^{(II)}}^{\mu}(x;{\mathbf{p}},\sigma) = \frac{-q {c_{-}^{(21;II)}}^{2}}{12 K^2{c_{-}^{(12;II)}}^{2}} [j^{(12;II) \mu}(x;{\mathbf{p}},\sigma) + j^{(21;II) \mu}(x;{\mathbf{p}},\sigma)] \quad ,
\end{equation}
\begin{equation} \label{CVP16}
  {a^{(III)}}^{\mu}(x;{\mathbf{p}},+1/2) = \frac{-q {c_{+}^{(12;III)}}^{2}}{12 K^2{c_{+}^{(21;III)}}^{2}} [j^{(12;III) \mu}(x;{\mathbf{p}},+1/2) + j^{(21;III) \mu}(x;{\mathbf{p}},+1/2)] \quad ,
\end{equation}
\begin{equation} \label{CVP17}
  {a^{(III)}}^{\mu}(x;{\mathbf{p}},-1/2) = \frac{-q {c_{-}^{(21;III)}}^{2}}{12 K^2{c_{-}^{(12;III)}}^{2}} [j^{(12;III) \mu}(x;{\mathbf{p}},-1/2) + j^{(21;III) \mu}(x;{\mathbf{p}},-1/2)] \quad ,
\end{equation}
where the constant $q$ is introduced to put the following equations in a familiar form. For each $A,$ one of the two currents $j^{(12;A)}$ or $j^{(21;A)}$ is constant, so including it does not change the equations below in which only derivatives of ${a^{(A)}}^{\mu}$ appear.

By (\ref{CVP9}), (\ref{CVP10}) and (\ref{CVP14}) it follows that ${a^{(A)}}^{\mu}$ is a vector potential for the constant current source $q j^{(12;A)\mu}$ or $q j^{(21;A)\mu}$ because the vector potential satisfies the relevant Maxwell equation. In detail, by (\ref{CVP9})-(\ref{CVP17}), the various ${a^{(A)}}^{\mu}$s obey the following equations:
\begin{equation} \label{CVP18}
  \partial^{\tau} \partial_{\tau} {a^{(I)}}^{\mu}(x;{\mathbf{p}},\sigma) -\partial^{\mu} \partial_{\kappa} {a^{(I)}}^{\mu}(x;{\mathbf{p}},\sigma)= q  j^{(12;I)\mu}(x;{\mathbf{p}},\sigma) \quad ,
\end{equation}
\begin{equation} \label{CVP19}
  \partial^{\tau} \partial_{\tau} {a^{(II)}}^{\mu}(x;{\mathbf{p}},\sigma) -\partial^{\mu} \partial_{\kappa} {a^{(II)}}^{\mu}(x;{\mathbf{p}},\sigma)= q  j^{(21;II)\mu}(x;{\mathbf{p}},\sigma) \quad ,
\end{equation}
\begin{equation} \label{CVP20}
  \partial^{\tau} \partial_{\tau} {a^{(III)}}^{\mu}(x;{\mathbf{p}},+1/2) -\partial^{\mu} \partial_{\kappa} {a^{(III)}}^{\mu}(x;{\mathbf{p}},+1/2)= q  j^{(12;III)\mu}(x;{\mathbf{p}},+1/2) \quad ,
\end{equation}
\begin{equation} \label{CVP21}
  \partial^{\tau} \partial_{\tau} {a^{(III)}}^{\mu}(x;{\mathbf{p}},-1/2) -\partial^{\mu} \partial_{\kappa} {a^{(III)}}^{\mu}(x;{\mathbf{p}},-1/2)= q  j^{(21;III)\mu}(x;{\mathbf{p}},-1/2) \quad .
\end{equation}
Equations (\ref{CVP18})-(\ref{CVP21}) show that the ${a^{(A)}}^{\mu}$ are vector potentials because the ${a^{(A)}}^{\mu}$ satisfy the Maxwell equations that constrain the vector potentials.

To obtain expressions for the associated electromagnetic fields, define the quantities ${F^{(A)}}^{\mu \nu}$ by
\begin{equation} \label{F1}
  {F^{(A)}}^{\mu \nu} = {a^{(A)}}^{\mu, \nu} - {a^{(A)}}^{\nu, \mu} \quad ,
\end{equation}
where the commas denote partial differentiation,
\begin{equation} \label{F2}
  {a^{(A)}}^{\mu, \nu} = \eta^{\nu \sigma} \frac{\partial{{a^{(A)}}^{\mu}}}{\partial{x^{\sigma}}} \quad .
\end{equation}
One of the Maxwell equations is satisfied directly by definition (\ref{F1}) since successive partials of $a^{\mu}$ commute,
\begin{equation} \label{F3}
  {F^{(A)}}^{\mu \nu, \lambda} + {F^{(A)}}^{\nu \lambda, \mu} + {F^{(A)}}^{\lambda \mu, \nu} = 0 \quad .
\end{equation}
By (\ref{CVP3}), (\ref{CVP4}) and (\ref{CVP14})-(\ref{CVP17}) one finds that
\begin{equation} \label{F4}
  {F^{(I)}}^{\mu \nu}(x;{\mathbf{p}},\sigma) = \frac{-q}{3}[ x^{\mu} j^{(12;I) \nu}(0;{\mathbf{p}},\sigma) -  x^{\nu} j^{(12;I) \mu}(0;{\mathbf{p}},\sigma)] \quad , \end{equation}
\begin{equation} \label{F5}
  {F^{(II)}}^{\mu \nu}(x;{\mathbf{p}},\sigma) = \frac{-q}{3}[ x^{\mu} j^{(21;II) \nu}(0;{\mathbf{p}},\sigma) -  x^{\nu} j^{(21;II) \mu}(0;{\mathbf{p}},\sigma)] \quad , \end{equation}
\begin{equation} \label{F6}
  {F^{(III)}}^{\mu \nu}(x;{\mathbf{p}},+1/2) = \frac{-q}{3}[ x^{\mu} j^{(12;III) \nu}(0;{\mathbf{p}},+1/2) -  x^{\nu} j^{(12;III) \mu}(0;{\mathbf{p}},+1/2)] \quad , \end{equation}
\begin{equation} \label{F7}
  {F^{(III)}}^{\mu \nu}(x;{\mathbf{p}},-1/2) = \frac{-q}{3}[ x^{\mu} j^{(21;III) \nu}(0;{\mathbf{p}},-1/2) -  x^{\nu} j^{(21;III) \mu}(0;{\mathbf{p}},-1/2)] \quad . \end{equation}
These equations imply that
\begin{equation} \label{F8}
 \frac{\partial{{F^{(I)}}^{\mu \nu}(x;{\mathbf{p}},\sigma)}}{\partial{\nu}} = q j^{(12;I) \mu}(x;{\mathbf{p}},\sigma) \quad , 
\end{equation}
\begin{equation} \label{F9}
 \frac{\partial{{F^{(II)}}^{\mu \nu}(x;{\mathbf{p}},\sigma)}}{\partial{\nu}} = q j^{(21;II) \mu}(x;{\mathbf{p}},\sigma) \quad , 
\end{equation}
\begin{equation} \label{F10}
 \frac{\partial{{F^{(III)}}^{\mu \nu}(x;{\mathbf{p}},+1/2)}}{\partial{\nu}} = q j^{(12;III) \mu}(x;{\mathbf{p}},+1/2) \quad , 
\end{equation}
\begin{equation} \label{F11}
 \frac{\partial{{F^{(III)}}^{\mu \nu}(x;{\mathbf{p}},-1/2)}}{\partial{\nu}} = q j^{(21;III) \mu}(x;{\mathbf{p}},-1/2) \quad . 
\end{equation}
Equations (\ref{F3}) and (\ref{F8})-(\ref{F11}) show that the ${F^{(A)}}^{\mu \nu}$ satisfy the Maxwell equations for the various source currents. Other electromagnetic fields can have the same charge current density; they differ from ${F^{(A)}}^{\mu \nu}$ by what are called `boundary conditions.'  The Invariant Coefficient Principle determines the fields ${F^{(A)}}^{\mu \nu}$ that satisfies the Maxwell equations for a particular set of boundary conditions.

It is important to emphasize that the current is quadratic in coefficient function factors giving rise to interference terms when coefficient functions are summed. Furthermore, the specific properties of the canonical vectors $a({\mathbf{p}},\sigma) $ may be relevant when coefficient functions for different momenta are mixed. See paper I, Appendix B, Problem 8. Since the discussions in this paper are limited to the consequences of the transformation properties of the canonical vectors, considerations based on the properties of annihilation operators lie beyond the scope of this paper and may be treated elsewhere. 

The translation-matrix-invariant coefficient functions may therefore be considered `intrinsically charged,' meaning the charges arises from the Invariant Coefficient Hypothesis and the transformation properties of the spacetime symmetry group connected to the identity. Thus intrinsic charge occurs not only for the massive spin 1/2 fields discussed in an earlier article, but also for the massless spin 1/2 fields  discussed in this article. 

\vspace{0.5cm}
%\pagebreak

\appendix

\section{Canonical Adjoint Representations} \label{adjoint}

The canonical fields $a({\mathbf{p}},\sigma)$ transform by a  representation of the Poincar\'{e} group, see equation (\ref{Da}) in the text. The Hermitian adjoint matrices also represent the group: if $D(W_1)D(W_2)$ = $D(W_1 W_2)$ then $D^{\dagger}(W_1)D^{\dagger}(W_2)$ = $(D(W_2)D(W_1))^{\dagger}$ = $D^{\dagger}(W_1 W_2).$ Thus the construction of covariant field vectors $\psi^{(12)}_{l}(x)$ and $\psi^{(21)}_{l}(x)$ could equally well proceed with vectors that transform by the adjoint representation.  

The canonical field vectors are denoted $a^{(c;B) \dagger}({\mathbf{p}},\sigma),$ with the representations labeled by $A \in$ $\{I,II,III\}$ in the text have adjoint representations $B \in$ $\{IV,V,VI\}.$  The superscript $c$ is a reminder that these may not be the adjoints of and may be entirely distinct from the canonical field vectors $a^{(A)}({\mathbf{p}},\sigma)$ used in the constructions in the text. A covariant field vector $\psi^{(B)}_{l}(x)$ constructed from the $a^{(c;B) \dagger}({\mathbf{p}},\sigma)$ can be termed a `negative energy field' while those of the text are `positive energy fields'. 

The constructions and derivations proceed much as in the text, so the presentation in this Appendix is abbreviated.

{\it{Invariant Coefficient Hypothesis}}. The covariant vector fields $\psi^{(12;B)}$ and $\psi^{(21;B)},$  are required to be  linear combinations of canonical field vectors $a^{(c;B) \dagger},$  
\begin{equation} \label{psi0A}
\psi^{(12;B)}_{l}(x) = \sum_{\sigma} \int d^3 p \enspace v^{(12;B)}_{l}(x;{\mathbf{p}},\sigma) a^{(c;B) \dagger}({\mathbf{p}},\sigma)  \quad ,
\end{equation}
and 
\begin{equation} \label{psi0A21}
\psi^{(21;B)}_{l}(x) = \sum_{\sigma} \int d^3 p \enspace v^{(21;B)}_{l}(x;{\mathbf{p}},\sigma) a^{(c;B) \dagger}({\mathbf{p}},\sigma)  \quad .
\end{equation}
The coefficient functions are now labeled $v.$

 The covariant fields $\psi^{(12;B)}_{l}(x)$ and $\psi^{(21;B)}_{l}(x)$ transform as before by (\ref{Dpsi2}) and (\ref{Dpsi2a}). The canonical field vectors $a^{(c;B)\dagger}$ transform via the adjoint to (\ref{Da}), 
\begin{equation} \label{DaA}
U(\Lambda,b) a^{(c;B)\dagger}({\mathbf{p}},\sigma) {U}^{-1}(\Lambda,b) = e^{i \Lambda p \cdot b} \sqrt{\frac{(\Lambda p)^t}{p^t}}  \sum_{\bar{\sigma}} D^{({\mathrm{Can}};B)\ast}_{\sigma \bar{\sigma}}(W^{-1}(\Lambda,p))  a^{(c;B)\dagger}({\mathbf{ p_{\Lambda}}},\bar{\sigma}) \quad ,
\end{equation}
where the matrix $D^{({\mathrm{Can}};B)\ast}(W^{-1}(\Lambda,p))$ belongs to representation $B.$ The representations $B$ are adjoints of the representations $A$ of the text.

The dependence of $v^{(12;B)}_{l}(x;{\mathbf{p}},\sigma)$ on coordinates $x$ and translation $b$ can be taken assimilated by defining $v_{\bar{l}}^{(B)}({\mathbf{p}},\sigma)$ in
\begin{equation} \label{Du2aA}
  v^{(12;B)}_{l}(x;{\mathbf{p}},\sigma) = (2 \pi)^{-3/2} e^{i p \cdot x} \sum_{\bar{l}} D^{(12)}_{l \bar{l}}(1,x)  v^{(B)}_{\bar{l}}({\mathbf{p}},\sigma) \quad 
\end{equation}
and 
\begin{equation} \label{Du2aA21}
  v^{(21;B)}_{l}(x;{\mathbf{p}},\sigma) = (2 \pi)^{-3/2} e^{i p \cdot x} \sum_{\bar{l}} D^{(21)}_{l \bar{l}}(1,x)  v^{(B)}_{\bar{l}}({\mathbf{p}},\sigma) \quad .
\end{equation}

The function $v^{(B)}_{\bar{l}}({\mathbf{p}},\sigma)$ may be different for the 12- and 21- representations, but it satisfies the same equation in both representations,
\begin{equation} \label{Du2cA}
    \sum_{\bar{l}} D_{l \bar{l}}(\Lambda) v^{(B)}_{\bar{l}}({\mathbf{p}},\sigma) = \sqrt{\frac{ (\Lambda p)^t}{p^t}} \sum_{\bar{\sigma}} v^{(B)}_{l}({\mathbf{ p_{\Lambda}}},\bar{\sigma}) D^{({\mathrm{Can}};B)\ast}_{\bar{\sigma} \sigma}(W(\Lambda,p))    \quad .
\end{equation}
One finds that the coefficient functions $v^{(12;B)}_{l}(x;{\mathbf{p}},\sigma)$ and $v^{(21;B)}_{l}(x;{\mathbf{p}},\sigma)$ are given by 
\begin{equation} \label{Du2hA}
  v^{(12;B)}_{l}(x;{\mathbf{p}},\sigma) = D^{(12)}_{l \bar{l}}(1,x) v^{(B)}_{\bar{l}}(x;{\mathbf{p}},\sigma) \quad \end{equation}
 and
\begin{equation} \label{Du2h1A}
v^{(21;B)}_{l}(x;{\mathbf{p}},\sigma) = D^{(21)}_{l \bar{l}}(1,x) v^{(B)}_{\bar{l}}(x;{\mathbf{p}},\sigma) \quad ,
\end{equation}
where $v^{(B)}(x;{\mathbf{p}},\sigma)$ is given by
\begin{equation} \label{Du2iA}
  v^{(B)}_{\bar{l}}(x;{\mathbf{p}},\sigma) = (2 \pi)^{-3/2}\sqrt{\frac{ k}{p^t}} e^{-i p \cdot x} \sum_{n} D_{\bar{l}n}(L(p))  v^{(B)}_{n}({\mathbf{k}},\sigma) \quad ,
\end{equation}
for both the 12- and 21-representations. 

It turns out that the generators for the adjoint of representation $I$ are similar to the generators of representation $II$ and visa versa. One can show by (\ref{jGENI}) and (\ref{jGENII}) that
\begin{equation} \label{genI}
  \{ \lambda_{1}, \lambda_{2},j \}^{(IV)} \equiv  -{\{\lambda_{1}, \lambda_{2}, j \}^{(I)}}^{\ast} = \sigma^{2}\{ \lambda_{1}, \lambda_{2},j \}^{(II)}  \sigma^{2} \quad ,
\end{equation}
where the notation is meant to imply that the factor of minus one, the complex conjugate indicated by the asterisk and the matrices $\sigma^{2}$ act on each generator in curly brackets. As discussed in the text, the $II$-generators $\{ \lambda_{1}, \lambda_{2},j \}^{(II)}$ are similar to the 22-block generators of the covariant transformation $\{L_{1}^{(22)},L_{2}^{(22)},J^{(22)}\}$ and are not similar to the 11-block generators. So the $IV$-generators are similar to the spin $(1/2,0)$ left-handed 22-block of covariant generators and not the spin $(1/2,0)$ right-handed 11-block generators. See Table 1. 

By deriving an adjoint version of (\ref{Du2d}) and then by Schur's Lemma, it follows that the right-handed upper two components of $v^{(IV)}$ must vanish. Then one finds that  $v^{(IV)}({\mathbf{k}},\sigma)$ is given by 
\begin{equation} \label{genIa}
  v^{(IV)}({\mathbf{k}},+1/2) = \pmatrix{0 \cr 0 \cr 0 \cr d_{-}^{(IV)} } \quad {\mathrm{and}} \quad v^{(IV)}({\mathbf{k}},-1/2) = \pmatrix{0  \cr 0 \cr - d_{-}^{(IV)} \cr 0 }   \quad .
\end{equation}

Let representation $V$ be the adjoint to representation $II.$ One finds that
\begin{equation} \label{genII}
  \{ \lambda_{1}, \lambda_{2},j \}^{(V)} \equiv  -{\{\lambda_{1}, \lambda_{2}, j \}^{(II)}}^{\ast} = \sigma^{2}\{ \lambda_{1}, \lambda_{2},j \}^{(I)}  \sigma^{2} \quad .
\end{equation}
Arguing as above for $IV,$ since the $I$-generators and, therefore, the $V$-generators are similar to the spin $(0,1/2)$ right-handed 11-block generators of the covariant transformation, it follows that the left-handed lower two components of $v^{(V)}$ must vanish, 
\begin{equation} \label{genIIa}
  v^{(V)}({\mathbf{k}},+1/2) = -\pmatrix{ 0 \cr d_{+}^{(V)} \cr 0 \cr 0} \quad {\mathrm{and}} \quad v^{(V)}({\mathbf{k}},-1/2) = \pmatrix{ d_{+}^{(V)} \cr 0 \cr 0 \cr 0}   \quad .
\end{equation}

Call the adjoint to the $III$-representation the $VI$-representation. One finds that
\begin{equation} \label{genIII}
  \{ \lambda_{1}, \lambda_{2},j \}^{(VI)} \equiv  -{\{0, 0, j \}^{(III)}}^{\ast} = \sigma^{2}\{ 0, 0,j \}^{(III)}  \sigma^{2} \quad .
\end{equation}
The $III$ and $VI$ representations do not have the equivalence problem of the others; the nonzero generator $j^{(III)}$ and $\sigma^{2} j^{(III)} \sigma^{2}$ are similar to the $J^{3}$ generator for both the 11-block and the 22-block. But to be equivalent to both blocks the canonical vectors must be eigenvectors of the $\lambda^{(I)}$ and $\lambda^{(II)}$ with eigenvalue zero, as previously discussed in the text in Section \ref{derive}. By similar reasoning as above, one finds that
\begin{equation} \label{genVIa}
  v^{(VI)}({\mathbf{k}},+1/2) = \pmatrix{0 \cr 0 \cr 0  \cr d_{-}^{(VI)} } \quad {\mathrm{and}} \quad v^{(VI)}({\mathbf{k}},-1/2) = -\pmatrix{ d_{+}^{(V)} \cr 0 \cr 0 \cr 0}   \quad .
\end{equation}
This completes the list of  quantities $v^{(B)}({\mathbf{k}},\sigma).$ 

Comparing the various $v^{(B)}({\mathbf{k}},\sigma)$ with the $u^{(A)}({\mathbf{k}},-\sigma)$ in Section \ref{derive} and by adjusting the constants $c$ and $d,$ it follows that
\begin{equation} \label{uvAB1}
  v^{(B)}({\mathbf{k}},\sigma)  = (2 \sigma) u^{(A)}({\mathbf{k}},-\sigma) \quad ,
\end{equation}
where $A$ and $B$ are paired as follows $\{(A,B)\}$ = $\{(II,IV),(I,V),(III,VI)\}.$ By  (\ref{Du2i}) and (\ref{Du2iA}), with the same pairing, one finds that
\begin{equation} \label{uvAB2}
  v^{(B)}(x;{\mathbf{p}},\sigma)  = (2 \sigma) u^{(A)}(-x;{\mathbf{p}},-\sigma) \quad ,
\end{equation}
where the $-x$ changes the phase $ip \cdot x$ for the $u^{(A)}$s to $-ip \cdot x$ for the $v^{(B)}$s, i.e. changing positive energy phase $-ip^{t}t$ to negative energy phase $+ip^{t}t.$ 
Finally, by (\ref{Du2a}), (\ref{Du2a21}), (\ref{Du2hA}),  and (\ref{Du2h1A}), it follows that
\begin{equation} \label{uvAB3}
  v^{(12;B)}_{l}(x;{\mathbf{p}},\sigma) = (2 \sigma) D^{(12)}_{l \bar{l}}(1,x)  u^{(12;A)}_{\bar{l}}(-x;{\mathbf{p}},-\sigma) \quad \end{equation}
and
\begin{equation} \label{uvAB4}
v^{(21;B)}_{l}(x;{\mathbf{p}},\sigma) = (2 \sigma) D^{(21)}_{l \bar{l}}(1,x)  u^{(21;A)}_{\bar{l}}(-x;{\mathbf{p}},-\sigma) \quad .
\end{equation}
Note that $(2 \sigma)$ is just a sign factor, plus or minus one, depending on whether $\sigma$ is $+ 1/2$ or $-1/2.$ Also note that the $\sigma$s switch $\sigma \rightarrow$ $-\sigma$ going from the left-hand side to the right-hand side above. Thus the adjoint (`negative energy') representations $B \in $  $\{I^{\dagger} = IV, II^{\dagger} = V, III^{\dagger} = VI\}$ are closely related to the positive energy representations in the text $A \in $  $\{II, I, III\},$ respectively. Simply stated, $I$ and $II$ are each other's adjoint while $III$ is its own adjoint.

A field more general than either a positive energy field discussed in the text or the negative energy fields of this Appendix is a linear combination of a positive and a negative energy field. But the fields must have the same transformation character to be invariantly combined. The fields $\psi^{(B)}_{\bar{l}}(x;{\mathbf{p}},\sigma)$ here and the $\psi^{(A)}_{\bar{l}}(x;{\mathbf{p}},\sigma)$ in the text can be paired so that they transform as covariant field vectors in the same way. Thus the positive energy field $\psi^{(I)}_{\bar{l}}(x;{\mathbf{p}},\sigma)$ with type $I$ transformation character can be combined with the negative energy field $\psi^{(V)}_{\bar{l}}(x;{\mathbf{p}},\sigma).$

{\it{Translation Matrix Invariance; Wave Equations}}. The translation matrices are in the form $1 - ix_{\mu} P^{\mu},$ so, when $P^{\mu} v^{(B)}$ = 0, the coefficient function $v^{(B)}$ is translation matrix invariant. By inspection of the column matrices above for the $v^{(B)}({\mathbf{k}},\sigma),$ one sees that $v^{(12;V)}_{l}(x;{\mathbf{p}},\sigma)$ and 
$v^{(12;VI)}_{l}(x;{\mathbf{p}},-1/2)$ are 12-translation matrix invariant because their lower two components are zero. Also $v^{(21;IV)}_{l}(x;{\mathbf{p}},\sigma)$ and $v^{(21;VI)}_{l}(x;{\mathbf{p}},+1/2)$ are 21-translation matrix invariant because their lower two components are zero.

Deriving wave equations for these translation matrix invariant coefficients functions is quickly done because wave equations were obtained in Section \ref{Dirac} for the associated $u^{(12;A)}$s and $u^{(21;A)}$s. For example, one has for $v^{(21;IV)}_{l}(x;{\mathbf{p}},-1/2),$
 \begin{equation} \label{aWE1}
i \delta^{\mu \nu}  \gamma_{\mu} \partial_{\nu} v^{(21;IV)}_{l}(x;{\mathbf{p}},-1/2) = (2 \sigma) i \delta^{\mu \nu}  \gamma_{\mu} \partial_{\nu} D^{(21)}_{l \bar{l}}(1,x)  u^{(21;II)}_{\bar{l}}(-x;{\mathbf{p}},+1/2) \quad 
\end{equation}
$$ \hspace{5cm} = -(-2) i \delta^{\mu \nu}  \gamma_{\mu} \partial_{\nu} u^{(21;II)}_{\bar{l}}(+x;{\mathbf{p}},+1/2) = 0 \quad . $$
Thus it can be shown based on (\ref{uvAB3}) and (\ref{uvAB4}) and the wave equations in Section \ref{Dirac}  that the translation matrix invariant fields satisfy the following wave equations
\begin{equation} \label{WE7A}
  i \gamma^{\mu} \partial_{\mu} \psi^{(21;IV)}(x) \mid_{\sigma = +1/2} = 0 \quad ,\end{equation}
\begin{equation} \label{WE8A}
 i \gamma^{\mu} \partial_{\mu} \psi^{(12;V)}(x) \mid_{\sigma = -1/2} = 0  \quad ,\end{equation}
\begin{equation} \label{WE9A}
 i \gamma^{\mu} \partial_{\mu} \psi^{(21;VI)}(x) \mid_{\sigma = +1/2} = 0  \quad ,\end{equation}
\begin{equation} \label{WE10A}
 i \gamma^{\mu} \partial_{\mu} \psi^{(12;VI)}(x) \mid_{\sigma = -1/2} = 0  \quad ,\end{equation}
\begin{equation} \label{PWE7A}
i \delta^{\mu \nu}  \gamma_{\mu} \partial_{\nu} \psi^{(21;IV)}(x) \mid_{\sigma = -1/2} = 0  \quad ,\end{equation}
\begin{equation} \label{PWE8A}
 i \delta^{\mu \nu}  \gamma_{\mu} \partial_{\nu} \psi^{(12;V)}(x) \mid_{\sigma = +1/2} = 0  \quad .\end{equation}
The first four of these has the scalar product using the spacetime metric $\eta_{\mu\nu},$ while the last two do not. These last two cannot be written as a covariant operator applied to $\psi.$ See the discussion in Section \ref{Dirac} and problems 5 and 6 in Appendix B.

{\it{Current as Vector Potential; the Maxwell Equations}}. The calculation showing that the current obeys the Maxwell equations goes through here just as in the text and in paper I. The reason is that the result follows from the coordinate dependence of the quantities ${D}^{(12)\dagger}(1,x) \gamma^{4}\gamma^{\mu} D^{(12)}(1,x)$ and ${D}^{(21)\dagger}(1,x) \gamma^{4}\gamma^{\mu} D^{(21)}(1,x)$ which is exactly the same as arose with massive particles in paper I. 

Define the currents $j^{(12;B) \mu}$ and $j^{(21;B) \mu}$ by 
\begin{equation} \label{CVP1A}
  j^{(12;B) \mu}(x;{\mathbf{p}},\sigma) = p^{t} \bar{v}^{(12;B)}(x;{\mathbf{p}},\sigma)\gamma^{\mu} v^{(12;B)}(x;{\mathbf{p}},\sigma) \quad 
\end{equation}
and
\begin{equation} \label{CVP2A}
  j^{(21;B) \mu}(x;{\mathbf{p}},\sigma) = p^{t} \bar{v}^{(21;B)}(x;{\mathbf{p}},\sigma)\gamma^{\mu} v^{(21;B)}(x;{\mathbf{p}},\sigma) \quad ,
\end{equation}
where $\bar{v}$ = $v^{\dagger} \gamma^{4}.$
These currents are constant in space and time when the coefficient function is translation-matrix-{\it{invariant}}, and depend on position when the coefficient function is translation-matrix-{\it{dependent}}. The translation-matrix-dependent currents are the vector potentials of the position independent currents.

As in the text, define the quantity ${a^{(B)}}^{\mu}$ to be proportional to the sum of the currents, 
\begin{equation} \label{CVP14A}
  {a^{(IV)}}^{\mu}(x;{\mathbf{p}},\sigma) = \frac{-q {d_{+}^{(12;IV)}}^{2}}{12 K^2{d_{+}^{(21;IV)}}^{2}} [j^{(12;IV) \mu}(x;{\mathbf{p}},\sigma) + j^{(21;IV) \mu}(x;{\mathbf{p}},\sigma)] \quad ,
\end{equation}
\begin{equation} \label{CVP15A}
  {a^{(V)}}^{\mu}(x;{\mathbf{p}},\sigma) = \frac{-q {d_{-}^{(21;V)}}^{2}}{12 K^2{d_{-}^{(12;V)}}^{2}} [j^{(12;II) \mu}(x;{\mathbf{p}},\sigma) + j^{(21;V) \mu}(x;{\mathbf{p}},\sigma)] \quad ,
\end{equation}
\begin{equation} \label{CVP16A}
  {a^{(VI)}}^{\mu}(x;{\mathbf{p}},+1/2) = \frac{-q {d_{+}^{(12;VI)}}^{2}}{12 K^2{d_{+}^{(21;VI)}}^{2}} [j^{(12;VI) \mu}(x;{\mathbf{p}},+1/2) + j^{(21;VI) \mu}(x;{\mathbf{p}},+1/2)] \quad ,
\end{equation}
\begin{equation} \label{CVP17A}
  {a^{(VI)}}^{\mu}(x;{\mathbf{p}},-1/2) = \frac{-q {d_{-}^{(21;VI)}}^{2}}{12 K^2{d_{-}^{(12;VI)}}^{2}} [j^{(12;VI) \mu}(x;{\mathbf{p}},-1/2) + j^{(21;VI) \mu}(x;{\mathbf{p}},-1/2)] \quad ,
\end{equation}
where the constant $q$ is introduced to put the following equations in a familiar form.

Again it follows that the $a^{(B)\mu}$ are vector potentials for the constant currents because one can show that they satisfy the Maxwell equations. For example,
\begin{equation} \label{CVP15A1}
  \partial^{\tau} \partial_{\tau} a^{(IV)\mu}(x;{\mathbf{p}},\sigma) -\partial^{\mu} \partial_{\kappa} a^{(IV)\kappa}(x;{\mathbf{p}},\sigma)= q  j^{(21;IV)\mu}(x;{\mathbf{p}},\sigma) \quad .
\end{equation}
The other Maxwell equations are similar; all have the same differential operator acting on $a^{(B)\mu}$ giving $q$ times the translation independent current which is either $j^{(12;B)}$ or $j^{(21;B)}.$ The electromagnetic field may be defined for each $B$ as previously in the text. The electromagnetic field satisfies the expected Maxwell equations, as may be quickly shown.

The coefficient functions $u^{12;A},$ $u^{21;A},$ $v^{12;B}$ and $v^{21;B}$ may therefore all be considered `intrinsically charged,' meaning the charge arises from the Invariant Coefficient Hypothesis and the transformation properties of the spacetime symmetry group connected to the identity. Having a charge is a consequence of the position dependence of the quantities ${D^{(12)}}^{\dagger}(1,x) \gamma^{4}\gamma^{\mu} D^{(12)}(1,x)$ and ${D^{(12)}}^{\dagger}(1,x) \gamma^{4}\gamma^{\mu} D^{(12)}(1,x)$ in (\ref{CVP3A}) and (\ref{CVP4A}), which arise in the construction of any current, whether it is the massive case or the massless case. Since charges occur automatically from the Invariant Coefficient Hypothesis for both the massive spin 1/2 particle case considered in paper I and the massless case considered in this article, the charge could be a `hypercharge' associated with a universal (at least for spin 1/2) `electromagnetic-like' interaction.

%\pagebreak

\section{Problems} \label{problems}

\noindent 1. (a) Use the matrices displayed in paper I(7) for the angular momentum generators $J^{\mu \nu}$ to verify the matrix expressions (\ref{L1}), (\ref{L2}) and (\ref{J}) for the generators of the little group $W,$ $\{L_1,L_2,J\}.$ (b) Show that these matrices satisfy the commutation rules in paper I among the generators $\{P^{x},P^{y},J^{12}\}.$
\vspace{0.3cm}

\noindent 2. With the regular representation of Poincar\'{e} transformations, i.e. spin $(1/2,1/2),$ the little group transformations $W$ preserve the 4-vector $k^{\mu}$ = $\{0,0,k,k\},$ with $k >$ 0. Thus $k^{\mu}$ is an eigenvector of $W$ with eigenvalue unity. (In general, for spin $(A,B)$ the eigenvalue would be $\exp{[i (A-B)  \theta ]},$ for the $\theta$ in (\ref{W}).) Find the eigenvectors of the 11-block of $D^{(12)}(W,0)$ = $D^{(21)}(W,0)$ = $D(W),$ and also of the 22-block, that have the appropriate eigenvalue $\exp{[i (A-B)  \theta ]}$ = $\exp{ (\pm i \theta /2)}.$ These are the `$k$-like vectors' of the spin 1/2 representation. [Hint: See paper I, Section 2 for the definitions of $D^{(12)}(\Lambda,b),$ $D^{(12)}(\Lambda,b),$ and $D(\Lambda).$ Can you get the results using the little group generators?] (b) Which coefficient functions $u_{l}^{(A)}({\mathbf{k}},\sigma)$ and $v_{l}^{(B)}({\mathbf{k}},\sigma)$ are $k$-like?
\vspace{0.3cm}

\noindent 3.   For the light-like, massless class of Poincar\'{e} transformations the momentum is a null 4-vector, $p_{\mu}p^{\mu}$ = 0. Show that this implies the second order wave equation 
\begin{equation} \label{PROB3}
\eta^{\mu \nu}\partial_{\mu}\partial_{\nu} \psi(x) = 0 \quad \end{equation}
for the following translation-matrix-invariant fields $\psi^{(12;I)}(x),$ $\psi^{(21;II)}(x),$ $\psi^{(12;III)}(x)\mid_{\sigma = +1/2},$ \linebreak $\psi^{(21;III)}(x)\mid_{\sigma = -1/2},$ $\psi^{(21;IV)}(x),$ $\psi^{(12;V)}(x),$ $\psi^{(21;VI)}(x)\mid_{\sigma = +1/2},$ and $\psi^{(12;VI)}(x)\mid_{\sigma = -1/2}.$
\vspace{0.3cm}

\noindent 4. (a) Show that the coefficient functions with transformation character $I$ are simply related to the coefficients with character $II.$ For example 
\begin{equation} \label{1221ePROB}
 \sqrt{p^{t}_{I}} u^{(12;I)}_{l}(\{{\mathbf{x}},x^{t}\};{\mathbf{p}}_{I},\sigma) = \gamma^4 \sqrt{p^{t}_{II}} u^{(21;II)}_{l}(\{-{\mathbf{x}},x^{t}\};{\mathbf{p}}_{II},\sigma)\mid_{c_{+}^{(I)} \leftrightarrow c_{-}^{(II)}}  \quad ,
\end{equation} 
where $p^{\mu}_{I}$ = $p^{\mu}(\xi,\theta,\phi)$ and $p^{\mu}_{II}$ = $p^{\mu}(-\xi,\theta,\phi).$ Note that ${\mathbf{p}}_{I}$ points in the same spatial direction as ${\mathbf{p}}_{II};$ see (\ref{WLp}) for the parameterization of $p^{\mu}.$
 (b) Compare $\gamma^{\mu} $ with $\gamma_{\mu}$ and $x^{\mu} $ with $x_{\mu},$ using the metric $\eta$ = diag$\{-1,-1,-1,+1\}.$ Can (\ref{1221ePROB}) be interpreted as a spacial inversion, i.e. a parity transformation? Why not? (c) Show that the same equation (\ref{1221ePROB}) holds for the massive class, except with momentum $p^{\mu}$ = $M\{\cos{\phi} \sin{\theta} \sinh{\xi},\sin{\phi} \sin{\theta} \sinh{\xi},$ $\cos{\theta} \sinh{\xi},\cosh{\xi}\}.$ Show that the massive class equation is easily seen to indicate a parity transformation. See paper I, equation I(57). (d) Discuss the massless class as a limit of the massive class as the mass $M$ goes to zero.

\vspace{0.3cm}

\noindent 5. Modify the expression  (\ref{mp1b}) with (\ref{mp1g}) for $\Pi(p_{\mu})$ to produce the following wave equations, 
\begin{equation} \label{PROB5i}
 i \gamma^{\mu} \partial_{\mu} \gamma^{4} \psi^{(12;I)}(x) \mid_{\sigma = -1/2} = 0  \quad ,\end{equation}
\begin{equation} \label{PROB5ii}
 i \gamma^{\mu} \partial_{\mu} \gamma^{4} \psi^{(21;II)}(x) \mid_{\sigma = +1/2} = 0  \quad .\end{equation}
These are clearly non-covariant because of the time component of the vector matrix, $\gamma^{4}.$
\vspace{0.3cm}

\noindent 6. If covariance is demanded but the differential equation requirement is dropped then another option opens up for the fields $\psi^{(12;I)}(x) \mid_{\sigma = -1/2}$ and $\psi^{(21;II)}(x) \mid_{\sigma = +1/2}.$ For example, show that $\Pi(p_{\mu})$ in (\ref{mp1b}) with (\ref{mp1g}) can be rewritten as 
\begin{equation} \label{PROB6i}
  \Pi(p_{\mu})  = \frac{{c_{+}^{(I)}}^2}{(2 \pi)^3}\frac{k^2}{{p^t}^2}\gamma^{4} p_{\mu} \gamma^{\mu} \gamma^{4} \quad .\end{equation}
Then with $p_{\mu}$ = $k_{\mu}$ = $\{0,0,-k,k\},$ by (\ref{vecTEN}), show that
\begin{equation} \label{PROB6ii}
 (p_{\mu} \gamma^{\mu} p^{(MASS)}_{\nu} \gamma^{\nu})_{l \bar{l}} u_{\bar{l}}^{(12;I)}(x;{\mathbf{p}},-1/2) = 0 \quad ,\end{equation}
\begin{equation} \label{PROB6iii}
 (p_{\mu} \gamma^{\mu} p^{(MASS)}_{\nu} \gamma^{\nu})_{l \bar{l}} u_{\bar{l}}^{(21;II)}(x;{\mathbf{p}},+1/2) = 0 \quad ,\end{equation}
where the contravariant components of $p^{(MASS)}$ given by
\begin{equation} \label{PROB6iv}
  {p^{(MASS)\mu}} = m\{ \cos{\phi} \sin{\theta} \sinh{\chi}, \sin{\phi} \sin{\theta} \sinh{\chi},\cos{\theta} \sinh{\chi},\cosh{\chi}\} \quad ,\end{equation}
and where $m \neq$ 0. Show that ${p^{(MASS)\mu}}$ obeys $p^{(MASS)}_{\mu}{p^{(MASS)\mu}}$ = $m^2.$ Thus ${p^{(MASS)\mu}}$ could be the momentum of a massive particle.
But the momentum $p^{(MASS)}$ is not a simple function of $p^{\mu}$ and so there is no corresponding simple differential wave equation for the fields $\psi^{(12;I)}(x) \mid_{\sigma = -1/2}$ and $\psi^{(21;II)}(x) \mid_{\sigma = +1/2}$ that follows from the identification of the light-like momentum with the differential operator $p_{\mu} \rightarrow$ $i\partial_{\mu}$. 
\vspace{0.3cm}

\noindent 7. (a) Use the formulas in the text to find explicit expressions for the electromagnetic fields ${F^{(A)}}^{\mu \nu}(x;{\mathbf{p}},\sigma)$ as a functions of position $x^{\mu}$ and energy $p^{t}$ = $k e^{\xi}$ for canonical transformation types $A \in$ $\{I,II,III\},$ of a charged massless spin 1/2 particle with momentum directed along the $z$-direction and with $\sigma$ = $+1/2.$ (b) Compare these results with the electromagnetic field of a massive spin 1/2 particle moving in the $z$-direction with an energy of $\cosh{\xi}$ times its rest energy, i.e. $p^t$ = $M \cosh{\xi},$ and with $\sigma$ = $+1/2.$ [See paper I, Problem 4(b) in Appendix B.]
\vspace{0.3cm}

%\pagebreak

\end{document}